\documentclass[nofootinbib]{revtex4}
\pdfoutput=1  
\usepackage[latin1]{inputenc}
\usepackage{ae,aecompl}
\usepackage{amsmath,amssymb,mathrsfs}
\usepackage{graphicx}
\usepackage[usenames,dvipsnames]{color} 
\usepackage{hyperref}
\hypersetup{
   colorlinks=true,       
    linkcolor=Blue,          
    citecolor=Plum,        
    filecolor=magenta,      
    urlcolor=YellowOrange           
}
\usepackage[]{units}

\providecommand{\hM}{\hat{M}}

\providecommand{\cp}{\mathsf{CP}}
\providecommand{\dcp}{\delta_{\cp}}

\providecommand{\om}{\omega}

\providecommand{\lag}{\mathcal{L}}

\providecommand{\tH}{\tilde{H}}
\providecommand{\ta}{\tilde{a}}
\providecommand{\tc}{\tilde{c}}
\providecommand{\bN}{\bar{N}}
\providecommand{\bS}{\bar{S}}
\providecommand{\bbM}{\mathbb{M}}

\providecommand{\eps}{\epsilon}
\providecommand{\nuless}{0\nu2\beta}

\providecommand{\Umutau}{\mathrm{U(1)}_{\mu-\tau}}
\providecommand{\cpmutau}{\cp^{\mu\tau}}
\providecommand{\md}[1]{m_{D_{#1#1}}}
\providecommand{\Msz}[1][]{M^{(0)}_{S_{#1}}}
\providecommand{\Ms}[1]{M_{S_{#1}}}
\providecommand{\cO}{\mathcal{O}}
\providecommand{\onel}{\text{1-l}}
\providecommand{\hX}{\hat{X}}
\providecommand{\hM}{\hat{M}}
\providecommand{\mbbeta}[1][]{m_{\beta\beta}^{#1}}
\providecommand{\pmns}{U_{\nu}}

\providecommand{\eqali}[1]{\begin{equation}\begin{aligned} #1
    \end{aligned}\end{equation}}
\providecommand{\eq}[1]{\begin{equation} #1 \end{equation}}
\providecommand{\mtrx}[1]{\begin{pmatrix} #1 \end{pmatrix}}
\DeclareMathOperator{\tr}{\mathrm{tr}} 
\providecommand{\bs}[1]{\boldsymbol{#1}}
\providecommand{\ZZ}{\mathbb{Z}}
\DeclareMathOperator{\diag}{\mathrm{diag}} 
\providecommand{\tp}{{\mss{\mathsf{T}}}}
\providecommand{\mss}[1]{\mbox{\scriptsize $#1$}}
\providecommand{\ml}[1]{\mbox{\large $#1$}}
\providecommand{\ums}[2][1]{\ml{\tfrac{#1}{#2}}} 
\providecommand{\aver}[1]{\langle #1 \rangle}
\usepackage{bbm}
\providecommand{\id}{{\mathbbm{1}}} 
\providecommand{\xlink}[1]
  {\href{http://arxiv.org/abs/#1}{#1}}
\providecommand{\doilink}[2]
  {\href{http://dx.doi.org/#1}{#2}}

\begin{document}
\title{
A new and trivial CP symmetry for extended $A_4$ flavor
}
\author{C.~C.~Nishi}
\email{celso.nishi@ufabc.edu.br}
\affiliation{
  Centro de Matemática, Computação e Cognição, Universidade Federal do ABC - UFABC, 
09.210-170,
Santo André, SP, Brasil
}
\begin{abstract}
The combination of $\nu_\mu$-$\nu_\tau$ exchange together with CP conjugation 
in the neutrino sector (known as $\cpmutau$ symmetry or $\mu\tau$-reflection) is 
known to predict the viable pattern: $\theta_{23}=45^\circ$, maximal Dirac CP 
phase and trivial Majorana phases.
We implement such a CP symmetry as a new CP symmetry in theories with $A_4$ flavor.
The implementation in a complete renormalizable model leads to a new form for the 
neutrino mass matrix that leads to further predictions: normal hierarchical 
spectrum with lightest mass and $\mbbeta$ ($\nuless$) of only few meV, and either 
$\nu_1$ or $\nu_2$ has opposite CP parity.
An approximate $L_\mu-L_\tau$ symmetry arises naturally and controls the flavor 
structure of the model. 
The light neutrino masses are generated by the extended seesaw mechanism with 6 
right-handed neutrinos (RHNs).
The requirement of negligible one-loop corrections to light neutrino masses, 
validity of the extended seesaw approximation and not too long-lived BSM states to 
comply with BBN essentially restricts the parameters of the model to a small region:
three relatively light right-handed neutrinos at the GeV-scale, heavier 
neutrinos at the electroweak scale and Yukawa couplings smaller than the electron 
Yukawa. Such a small Yukawa couplings render these RHNs unobservable in terrestrial 
experiments.

\end{abstract}
\maketitle
\section{Introduction}
\label{sec:intro}

The discovery of nonzero $\theta_{13}\sim 8.5^\circ$ in 2012\,\cite{theta13} 
prompted the neutrino physics community to one of its next experimental 
goals: measure or discard CP violation in the leptonic sector\,\cite{snowmass:nu}.
As one more parameter in the standard three neutrino paradigm joined the list of 
known quantities, we are only left with three unknowns in case neutrinos are 
Majorana: neutrino mass ordering, absolute neutrino mass scale and CP violation in 
the leptonic sector.
The last unknown has three sources: one Dirac CP phase analogous 
to the CKM phase for quarks and two Majorana phases.

From a theory viewpoint, many symmetries were sought over the years in 
order to predict the CP violating phases of the leptonic sector.
The simplest of them that leads to CP violation and viable mixing angles is known 
as $\mu\tau$-reflection or $\cpmutau$ which consists on $\nu_\mu$-$\nu_\tau$ flavor 
exchange together with CP conjugation\,\cite{mutau-r}.
Often, such a CP symmetry is considered in conjunction with nonabelian discrete 
symmetries\,\cite{s4-tilde,hagedorn,holthausen,ratz}.
In fact, many studies were devoted to the definition of CP symmetry in that 
context\,\cite{hagedorn,holthausen,ratz}.
However, differently from many simple flavor symmetries that predicted 
vanishing $\theta_{13}$, the $\cpmutau$ symmetry allows nonzero $\theta_{13}$ but 
predicts all the presently unknown CP phases: the Dirac CP phase $\dcp=\pm 
90^\circ$ is maximal while the Majorana phases are trivial\,\cite{mutau-r,cp.mutau}.
Moreover, $\theta_{23}$ is also predicted to be maximal, the neutrinoless double 
beta decay effective mass is restricted to narrower bands and, in simple 
implementations, leptogenesis is only allowed to occur in the intermediate range of 
$T\sim M_1\sim 10^9\text{ -- }10^{12}\,\unit{GeV}$ where flavor 
effects are important\,\cite{cp.mutau}.
From current global fits\,\cite{GG:fit,fit:others}, we know in fact there is a 
slight preference for negative $\dcp$ and $\theta_{23}=45^\circ$ is still allowed.

Two directions were recently pursued to generalize the idea of $\cpmutau$ symmetry. 
Firstly, we have 
shown in Ref.\,\cite{cp.mutau} that a minimal setting that allowed distinct 
symmetries in the charged lepton and neutrino sectors consisted of only one abelian 
symmetry (the combination of lepton flavors $\mathtt{L}_\mu-\mathtt{L}_\tau$ or 
subgroup) and CP symmetry ($\cpmutau$).
This setting was shown to be free from the vev alignment problem that plagues many 
flavor symmetry models for leptons.
In contrast, in Ref.\,\cite{real.sym}, it was shown that maximal $\theta_{23}$ and 
$\dcp$ (the prediction for Majorana phases is lost) could follow from much 
more general assumptions 
without the imposition of CP symmetry. The necessary conditions involve the 
symmetry of the charged lepton sector ($G_l)$ to be represented by \textit{real} 
matrices in the flavor space and, in the 
same basis, $M_\nu$ needs to be diagonalizable by a \textit{real} matrix.
The crucial aspect is the former, which presumably follows from a 
\textit{real} flavor symmetry 
conserved in the charged lepton sector. The neutrino sector cannot be invariant by 
the same residual symmetry and hence must have a large breaking in the form of 
misaligned vevs.

Here we try to embed a subgroup of $\mathtt{L}_\mu-\mathtt{L}_\tau$ into 
a discrete nonabelian flavor 
group $G_F$ in order to increase predictivity but, at the same time, retain
the successful features of Ref.\,\cite{cp.mutau}.
We choose the $A_4$ group which is an extensively studied flavor group (see 
\cite{review} and references therein).
In fact, the first $\cpmutau$ symmetric neutrino mass matrix was obtained 
with this group\,\cite{babu}.
More recent studies involving $A_4$ and CP can be seen in Refs.\,\cite{a4.cp,ding}.

We anticipate that the light neutrino mass matrix in our model will have the form
\eqali{
  \label{Mnu:a4cp}
M_\nu = \mtrx{a_1+a_2+a_3 & k(a_1+\om a_2+\om^2 a_3) & k(a_1+\om^2 a_2+\om a_3) 
\cr 
 k(a_1+\om a_2+\om^2 a_3) & k^2(a_1+\om^2 a_2+\om a_3) & k^2(a_1+a_2+a_3)
\cr
 k(a_1+\om^2 a_2+\om a_3) & k^2(a_1+a_2+a_3) & k^2(a_1+\om a_2+\om^2a_3) 
}\,,
}
where $a_i,k$ are real parameters and $k>0$ can be chosen;
$\om\equiv e^{i2\pi/3}$ as usual.
This mass matrix is $\cpmutau$ symmetric\,\cite{mutau-r} but has 4 real parameters 
to describe 5 observables: $\theta_{12},\theta_{13},m_1,m_2,m_3$. Hence, we will
have one prediction.

The paper is organized as follows: In Sec.\,\ref{sec:a4} we describe the new CP 
symmetry that can be implemented for theories with $A_4$ symmetry.
Section \ref{sec:mass} shows that the mass matrix \eqref{Mnu:a4cp} can fit the 
present oscillation parameters and additionally give predictions for the absolute 
neutrino mass and CP parities.
A complete renormalizable model is shown in Sec.\,\ref{sec:ESS} where the 
light neutrino masses are generated by the extended seesaw (ESS) 
mechanism\,\cite{ess} with relatively light right-handed neutrinos in its spectrum.
The approximate symmetry $\mathtt{L}_\mu-\mathtt{L}_\tau$ is presented in 
Sec.\,\ref{sec:Umutau} and shown to constrain the flavor structure of the model.
Section\,\ref{sec:one-loop} analyzes the constraints on the model coming from 
(i) the radiative stability of the tree-level result, (ii) validity of the ESS 
approximation to fit the light neutrino masses and (iii) sufficiently short-lived 
BSM states that not spoil Big Bang nucleosinthesis.
More phenomenological constraints on the presence of relatively light 
right-handed neutrinos are analyzed in Sec.\,\ref{sec:pheno}.
The conclusions are shown in Sec.\,\ref{sec:concl} and the appendices contain 
auxiliary material.

\section{Another GCP for $A_4$}
\label{sec:a4}

The group $A_4=(\ZZ_2\times\ZZ_2)\rtimes\ZZ_3$ has one three-dimensional 
irreducible representation (irrep) $\bs{3}$ and three one-dimensional irreps 
$\bs{1}',\bs{1}'',\bs{1}$, where the latter is the trivial invariant (singlet).
The faithful $\bs{3}$ can be generated by
\eq{
\label{irrep:3}
a=\diag(1,-1,-1),\quad
b=\mtrx{0&1&0\cr0&0&1\cr1&0&0}\,,
}
where $a$ generates one of the $\ZZ_2$ subgroups and $b$ generates the $\ZZ_3$ 
subgroup. Only $b$ acts nontrivially on the singlets $\bs{1}',\bs{1}''$ as
\eq{
\label{irrep:1}
\bs{1}'\stackrel{b}{\to} \om\bs{1}',\quad
\bs{1}''\stackrel{b}{\to} \om^2\bs{1}''\,,
}
where $\om=e^{i2\pi/3}$.

For generic settings where generic irreps of $A_4$ (e.g. a $\bs{3}$ and one
\textit{charged} $\bs{1}'$) are considered in a model, there is only one possible 
CP symmetry that can be imposed on the model\,\cite{holthausen,ratz}.
As first considered in Ref.\,\cite{s4-tilde},%
\footnote{
For the triplet $\bs{3}$ only, in the context of the $A_4$ invariant 
3HDM, it was first considered in Ref.\,\cite{toorop} (erratum) as an accidental 
symmetry and in Ref.\,\cite{igor} in the course of symmetry classification.
}
CP acts on the representations of 
\eqref{irrep:3} and \eqref{irrep:1} as
\eq{
  \label{gcp1}
\cp_1:\quad \bs{3}\to X\bs{3}^*,\quad
\bs{1}\to \bs{1}^*,\quad
\bs{1}'\to {\bs{1}'}^*,\quad
\bs{1}''\to {\bs{1}''}^*\,,
}
where $X$ can be chosen as $(23)$ exchange:
\eq{
X=\mtrx{1&0&0\cr 0&0&1\cr0&1&0}\,.
}
The complex conjugation denotes the CP transformation operation on the fields which 
should be adjoined with the appropriate Lorentz factors for e.g. spin 1/2 fermions.
We denote the whole flavor group considering $\cp_1$ as $A_4\rtimes\ZZ_2^\cp$ and it
gives rise to a group isomorphic to $S_4$, denoted as $\tilde{S}_4$ in 
\cite{s4-tilde}.
Obviously any composition of $\cp_1$ with an element of $A_4$ is also a GCP 
symmetry, so any of the 12 GCP symmetries can be chosen as a residual 
symmetry\,\cite{ding}.

In nongeneric settings where only a specific set of irreps is considered, it is 
clear that there is one more inequivalent option. If only $\bs{3}$ is considered, 
we can use the \textit{usual} CP transformation\,%
\footnote{It is important to note that the GCP \eqref{gcp1}, with symmetric $X$, 
can also be cast in the form \eqref{gcp2} by basis change, after which the 
representation \eqref{irrep:3} changes and is no longer manifestly real.}:
\eq{
\label{gcp2}
\cp_2:\bs{3}\to \bs{3}^*\,.
}
Given that the representation \eqref{irrep:3} is real, the whole group including 
$\cp_2$ will be denoted as $A_4\times\ZZ_2^\cp$ where $\ZZ_2^\cp$ is generated 
by $\cp_2$, which commutes with $A_4$ ($\bs{3}$ is real).

Now the question is: What is the transformation law for the other irreps (if 
any is consistent)?
We can deduce them by noting that the transformation \eqref{gcp2} acts on the 
representation \eqref{irrep:3} \textit{trivially}, i.e., 
\eq{
\label{auto:2}
\cp_2:\quad
a\to a,\quad
b\to b\,,
}
if we apply on any $\bs{3}$, in this order, $\cp_2$, the transformation $a$ or $b$ 
and then $\cp_2^{-1}$.
In contrast, for $\cp_1$, the same set of operations induces
\eq{
\label{auto:1}
\cp_1:\quad
a\to Xa^*X^{-1}=a,\quad
b\to Xb^*X^{-1}=b^2.
}
Here we are identifying $a,b$ with its three-dimensional irrep 
$D_{\bs{3}}(a),D_{\bs{3}}(b)$ in \eqref{irrep:3}.
Given that \eqref{auto:1} and \eqref{auto:2} lead to different rules (map different
conjugacy classes), they cannot be equivalent. These mapping rules in the group 
are called automorphisms and only \eqref{auto:1} and \eqref{auto:2} are 
nonequivalent for  $A_4$. So these are the only 
possibilities for defining GCP in the presence of $A_4$ 
symmetry\,\cite{holthausen}.'

We can now deduce that one transformation law for the singlets $\bs{1}'$ 
that is compatible with \eqref{gcp2} and \eqref{auto:2} is the \textit{trivial}
transformation
\eq{
  \label{cp2:neutral}
\cp_2:\quad
\bs{1}'\to \bs{1}'\,.
}
However, this transformation law can only be used if the complex field $\psi_1\sim 
\bs{1}'$ is neutral under any other group, including the Lorentz group, i.e., it 
must be a scalar\,\footnote{%
One could also use \eqref{cp2:neutral} as charge conjugation for a 
\textit{pair} of Majorana fermion fields where $b$ acts by $120^\circ$ rotation in 
the plane.
}.
In this case, we can split any complex scalar into its real and 
imaginary parts, $\varphi=(\varphi_r+i\varphi_i)/\sqrt{2}$, and consider the action 
of $b$ of $A_4$ as a $120^\circ$ rotation in the plane of 
$(\varphi_r,\varphi_i)^\tp$, hence a \textit{real} representation that is trivial 
under $\cp_2$, i.e., $\varphi_r,\varphi_i$, are CP-even real scalar fields.

On the other hand, if $\psi_1$ carries other complex quantum numbers (it excludes 
$\ZZ_2$) other than $A_4$, say a charge $q$ of $U(1)$, then \eqref{cp2:neutral} is 
not compatible with the fact that CP should reverse the charge $q$.
Therefore, in this case another field $\psi_2\sim\bs{1}''$ with the same charge $q$ 
(or any other quantum number) needs to be introduced to define the transformation 
\eq{
\cp_2:\quad \psi_1\to \psi_2^*\,,
}
so that both sides transform as $\om$ by $b$ but the field of charge $q$ is mapped 
to a field of charge $-q$.
This is also the transformation law for fermions.
To summarize, the irreps $\bs{1}'$ and $\bs{1}''$ are \textit{exchanged} by $\cp_2$,
\eq{
\label{cp2:charged}
\cp_2:\quad \bs{1}'\to {\bs{1}''}^*,
}
unless ${\bs{1}''}^*$ can be identified with $\bs{1}'$.
Therefore, for charged fields (such as the SM fields or any chiral fermion) the 
irreps $\bs{1}',\bs{1}''$ 
need to be introduced in \textit{pairs}.
It is always possible to recast \eqref{cp2:charged} as the usual CP transformation 
by changing basis; see appendix B of Ref.\,\cite{cp.mutau} for the explicit basis 
change.

Compatibility with the triplet transformation law \eqref{gcp2} can also be checked 
independently by forming an invariant with two triplets 
$\psi=(\psi_i),\chi=(\chi_i)\sim \bs{3}$ (say fermionic and left chiral) and a 
scalar $\varphi\sim \bs{1}'$, and ensuring that $\cp_2$ maps an $A_4$ invariant to 
an $A_4$ invariant\,\cite{holthausen}.
The only trilinear $A_4$ invariant involving $\bar{\psi},\chi$ and $\varphi$ is
\eq{
I=(\bar{\psi}_1\chi_1+\om\bar{\psi}_2\chi_2+\om^2\bar{\psi} _3\chi_3)\varphi\,.
}
It is tranformed by \eqref{gcp2} (for $\psi,\chi$) and \eqref{cp2:neutral} (for 
$\varphi$) to
\eq{
\cp_2:\quad
I\to 
(\bar{\chi}_1\psi_1+\om\bar{\chi}_2\psi_2+\om^2\bar{\chi}_3\psi_3)\varphi\,,
}
which remains as an $A_4$ invariant.

The symmetry $\cp_2$ (associated to the trivial automorphism) can be 
straightforwardly extended for other groups with structure $H\rtimes \ZZ_3$ such as 
the $\Delta(3\cdot N^2)=(\ZZ_N\times\ZZ_N)\rtimes \ZZ_3$ family [e.g. 
$\Delta(27)$\,\cite{D27}] or some of its subgroups such as $T_7$ or $T_{13}$. The 
only difference is that the triplet representations would be complex and CP 
symmetry would act as usual.

We stress that the $\cp_2$ symmetry for $A_4$ has not been considered for flavor 
model building before.
This possibility is raised in the general context of discrete nonabelian symmetries 
in \cite{holthausen} but no model application was discussed.
For $A_4$, this possibility was mentioned in \cite{ding} but it was not pursued.
Ref.\,\cite{ratz} discards this kind of CP symmetry dubbing it as CP-like 
symmetries but---as we will see for the simple case of $A_4$---no theoretical 
consideration prevents its use.
As an added bonus, we will see that the transformation property \eqref{cp2:neutral} 
allows us to avoid the vev alignment problem\,\cite{cp.mutau}.

\section{Mass matrix}
\label{sec:mass}

We first analyze our mass matrix \eqref{Mnu:a4cp} in the flavor basis to show that 
we can correctly fit the oscillation parameters.
This is a new form for the neutrino mass matrix that has not been considered so far.

The $\cpmutau$ symmetry of \eqref{Mnu:a4cp} implies that $\theta_{23}=\pi/4$ and 
$\dcp=\pm \pi/2$ are automatic\,\cite{mutau-r} and the diagonalization
\eq{
U^\tp M_\nu U=\diag(m_i')\,,
}
can be performed by a matrix $U=U_0$ of the form
\eq{
U_0=\mtrx{u_1&u_2&u_3\cr w_1& w_2 & w_3\cr w_1^*& w_2^* & w_3^*}\,,
}
with $u_i$ conventionally real and positive.
The Majorana phases are trivial and possible CP parities appear along with the 
eigenvalues $m_i'=\pm m_i$, $m_i\ge 0$.
We denote the different cases of CP parities by the sign of $(m'_i)$ as
\eq{
(+++), (-++), (+-+), (++-)\,.
}

In addition to being $\cpmutau$ symmetric, the mass matrix in 
\eqref{Mnu:a4cp} 
obeys
\eqali{
  \label{prop:Mnu}
M_\nu\big|_{a_2\leftrightarrow a_3}&=M_\nu^*\,,\cr
M_\nu\big|_{a_1\to a_2\to a_3}&=\diag(1,\om^2,\om)M_\nu\diag(1,\om^2,\om)\,.
}
Thus cyclic permutation of $a_i$ leaves all observables of $M_\nu$ invariant 
while a transposition ($a_2\leftrightarrow a_3$) flips the Dirac CP phase: $\dcp\to 
-\dcp$.
Hence, permutations of solutions for $a_i$ are solutions as well.

\subsection{Obtaining the masses}

To extract the light neutrino masses, it is more convenient to change to a 
real basis:
\eq{
M_\nu'=U_{23}^\tp M_\nu U_{23}=
\left(
\begin{array}{ccc}
 a_1+a_2+a_3 & \frac{k}{\sqrt{2}}(2 a_1-a_2-a_3)
  & \sqrt{\frac{3}{2}}k (a_3-a_2) \\
\frac{k}{\sqrt{2}}(2 a_1-a_2-a_3) & 
\frac{1}{2}k^2(4a_1+a_2+a_3) & \frac{1}{2}\sqrt{3}k^2(a_2-a_3) \\
 \sqrt{\frac{3}{2}}k(a_3-a_2) & \frac{1}{2}\sqrt{3}k^2(a_2-a_3) & 
  \frac{3}{2}k^2(a_2+a_3)
\end{array}
\right)
\,,
}
where
\eq{
\label{U23}
U_{23}\equiv
\left(
\begin{array}{ccc}
 1 & 0 & 0 \\
 0 & \frac{1}{\sqrt{2}} & \frac{i}{\sqrt{2}} \\
 0 & \frac{1}{\sqrt{2}} & -\frac{i}{\sqrt{2}} 
\end{array}
\right)\,.
}
Now $M_\nu'$ is real symmetric and can be diagonalized by a real orthogonal matrix.

The eigenvalues of $M_\nu'$ will correspond to the light neutrino masses $m_i'=\pm 
m_i$ with its CP parities.
They are solutions of the characteristic equation 
\eq{
  \label{cubic}
\lambda^3+c_1\lambda^2+c_2\lambda+c_3=0\,,
}
with coefficients
\eqali{
  \label{ai->mi}
-c_1&=(1+2k^2)(a_1+a_2+a_3)=m_1'+m_2'+m_3'\,\cr
-c_3&=27k^4a_1a_2a_3=m_1'm_2'm_3'\,\cr
c_2&=3k^2(2+k^2)(a_1a_2+a_2a_3+a_3a_1)
  =m_1'm_2'+m_2'm_3'+m_3'm_1'\,.
}

It is clear that $k=1$ is a special point where
\eq{
  \label{ai=mi}
3a_i=m_i'\,,\quad i=1,2,3\,,
}
is a solution; permutation of $a_i$ still leads to a solution.
However, our mass matrix \eqref{Mnu:a4cp} with $k=1$ and with the second and third 
columns (rows) exchanged is invariant by cyclic permutations which means it is 
diagonalized by $U_{\rm PMNS}=U_\om$.
This mixing matrix is clearly in contradiction with experiments, a fact that still 
applies if $k\approx 1$ (for hierarchical $m_i$).
Hence, we need to analyze the cases away from $k=1$.

Generically we can invert \eqref{ai->mi} and obtain $a_i$ as functions of $m_i$ and 
$k$.
A simplification is achieved for generic $k>0$ by defining
\eq{
\ta_i\equiv (1+2k^2)a_i\,.
}
Then the equations in \eqref{ai->mi} can be rewritten as
\eqali{
\label{ai->mi:2}
\ta_1+\ta_2+\ta_3&=m_1'+m_2'+m_3'\,,\cr
\ta_1\ta_2\ta_3&=g_3(k)m_1'm_2'm_3'\,,\cr
\ta_1\ta_2+\ta_2\ta_3+\ta_3\ta_1&=g_2(k)
(m_1'm_2'+m_2'm_3'+m_3'm_1')\,.
}
where
\eq{
g_3(k)\equiv \frac{(1+2k^2)^3}{27k^4}\,,\quad
g_2(k)\equiv\frac{(1+2k^2)^2}{3k^2(2+k^2)}\,.
}
The key relation that can be extracted from \eqref{ai->mi:2} is that $\ta_i$ should 
now be roots of the cubic equation similar to \eqref{cubic} 
but with coefficients modified by
\eq{
  \label{c-tilde}
c_1\to \tc_1=c_1\,,\quad
c_2\to \tc_2=g_2(k)c_2\,,\quad
c_3\to \tc_3=g_3(k)c_3\,.
}
This construction gives $\ta_i$ as functions of $m_i'$ and $k$, except for
permutations of $\ta_i$.
The solutions \eqref{ai=mi} for $k=1$ are modified as $g_2(k),g_3(k)$ differ from 
unity when $k\neq 1$. Moving away from $k=1$, both functions 
increase monotonically ($g_2$ reaches $4/3$ asymptotically as $k\to\infty$).

Now, the distortions caused by $g_{2,3}$ cannot be too large because the $\ta_i$ 
need to be real.
To illustrate this point, compare the two polynomials
\eq{
p_1(x)=x^3-2.1x^2+1.1x\,,\quad
p_2(x)=x^3-2.1x^2+1.2x\,,
}
where the second polynomial differs from the first just by a small deviation in the 
third coefficient.
The first polynomial has three real and distinct roots
while the second polynomial has only $x=0$ as a real root.
This can be confirmed by calculating the discriminant of the factored second-degree 
polynomials: $\Delta=(2.1)^2-4\times 1.1=0.01$ and $\Delta=(2.1)^2-4\times 
1.2=-0.39$ for $p_1$ and $p_2$ respectively.
We can see that two quasidegenerate eigenvalues are specially sensitive to 
deviations by $k$.
This is the case of IH with CP parities $(+++)$ or $(++-)$.

\begin{figure}[h]
\centering
\includegraphics[scale=0.41,angle=0]{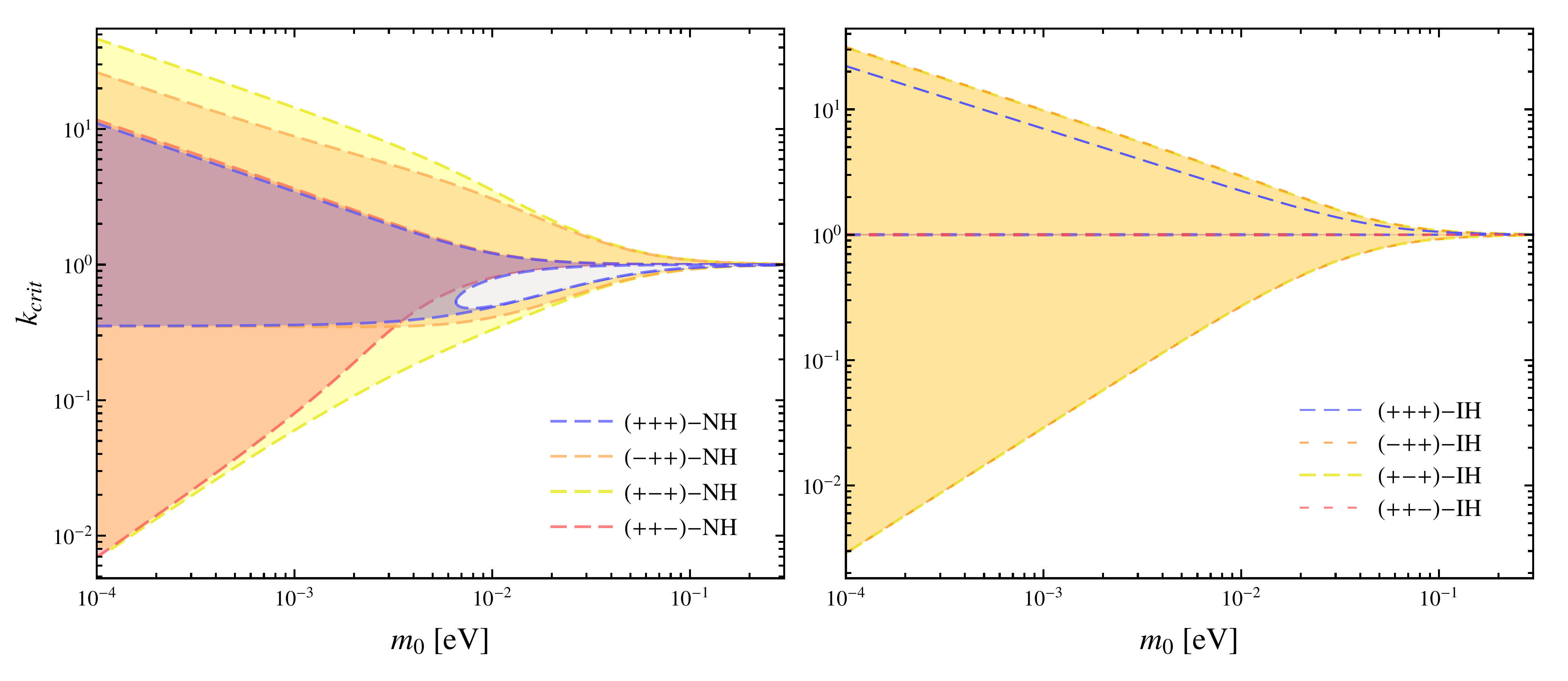}
\caption{
Left (right): Regions in the $k$-$m_1$ ($k$-$m_3$) plane where solutions for $a_i$ 
are real for NH (IH).
The mass-squared differences are fixed to their best-fit values of \cite{GG:fit}.
A hole is only present for the case NH-$(+++)$, the regions for IH-$(-++)$ and 
IH-$(+-+)$ are almost overlapping, and the regions for IH-$(+++)$ and IH-$(++-)$ 
can be seen only as lines.
}
\label{fig:kcrit}
\end{figure}

The values for $k$ that allow real solutions for $\ta_i$ can be extracted from the 
discriminant of the cubic polynomial \eqref{cubic} for which 
\eq{
  \label{disc}
\Delta=\tc_1^2\tc_2^2-4\tc_2^3-4\tc_1^3\tc_3+18\tc_1\tc_2\tc_3-27\tc_3^2~\ge 0\,.
}
In Fig.\,\ref{fig:kcrit} we show the values of $k$ as a function of the lightest 
mass $m_0$ where the discriminant above is non-negative; we use the current best 
fit values for the mass differences\,\cite{GG:fit}.
The figure on the left (right) corresponds to NH (IH) and the various possibilities 
for CP parities are depicted in different colors.
For IH, only the case of CP parities $(-++)$ and $(+-+)$ have wide regions for $k$ 
for a given mass $m_0$; the remaining cases only have very narrow ranges of 
possible $k$, including $k\approx 1$ which is phenomenologically excluded.
The other possible narrow range for $k$ for IH-$(+++)$ (e.g. $k\approx 7$ for 
$m_0=10^{-3}\rm eV$) is also phenomenologically excluded because it leads to 
$a_1\approx a_2\approx a_3$ and two mixing angles are vanishing.

\begin{figure}[h]
\centering
\includegraphics[scale=0.35,angle=0]{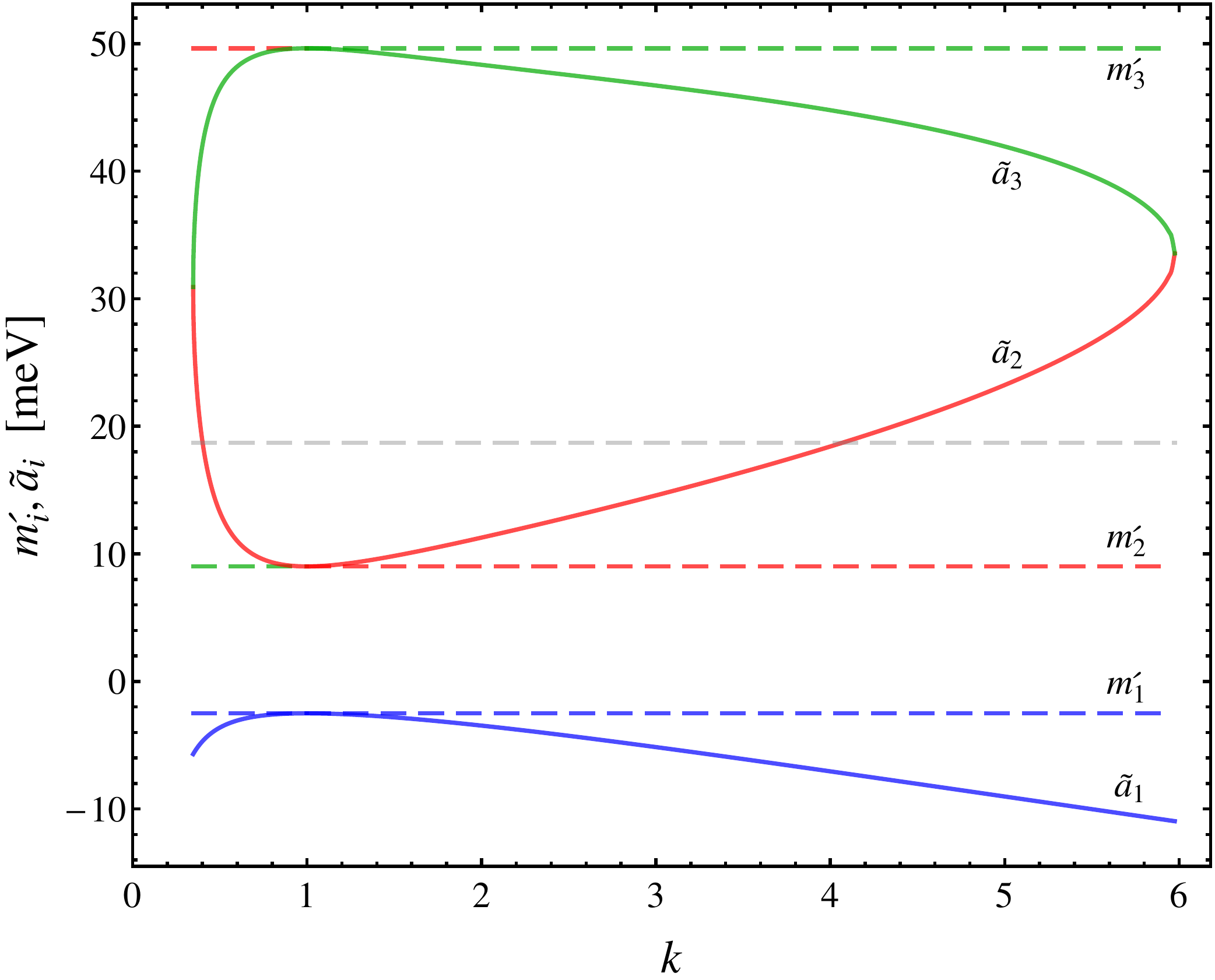}
\caption{
Solutions of $\ta_i$ (in solid blue, red and green, respectively) as functions of 
$k$ and $m'_i$ (in dashed blue, red and green, respectively) for fixed values 
$(-2.5,9.014,49.63)\rm\, meV$ for $m'_i$ with ordering defined by \eqref{order}. 
The gray dashed line corresponds to the average 
$\ums{3}(m_1'+m_2'+m_3')=\ums{3}(\ta_1+\ta_2+\ta_3)$.
}
\label{fig:miai}
\end{figure}

We also illustrate in Fig.\,\ref{fig:miai} the deviations from $\ta_i=m_i$ when $k$ 
moves away from $k=1$.
$k$ varies only in the range where the discriminant \eqref{disc} is 
non-negative, as shown in Fig.\,\ref{fig:kcrit}.
Note that close to the critical values of $k$ ($\Delta=0$) two (or more) $\ta_i$ 
tend to be quasidegenerate.
This is a generic phenomenon.

\subsection{Seeking solutions}
\label{sec:sols}

After an exhaustive numerical search we conclude that the mass matrix 
\eqref{Mnu:a4cp} is only compatible with oscillation data for normal hierarchy (NH) 
and CP parities $(-++)$ and $(+-+)$.
The cases of IH and quasidegenerate masses are excluded.
The lightest neutrino mass is restricted to
\eqali{
    \label{m1:range}
(-++):&& 1.6\,\text{meV}&\lesssim m_1\lesssim 3\,\text{meV}\,,\cr
(+-+):&& 3.5\,\text{meV}&\lesssim m_1\lesssim 7.7\,\text{meV}\,.
}
The predictions for the contribution for neutrinoless double-beta decay coming from 
light neutrinos is given by
\eqali{
  \label{nuless:0}
(-++):&& 1.9\,\text{meV}&\lesssim |\mbbeta[\nu]|\lesssim 2.6\,\text{meV}\,,\cr
(+-+):&& 1.1\,\text{meV}&\lesssim |\mbbeta[\nu]|\lesssim 2.05\,\text{meV}\,,
}
They fall inside the regions denoted by NH-$(-++)$ and NH-$(+-+)$ in 
Ref.\,\cite{cp.mutau}.
Note that $\mbbeta[\nu]=(M_\nu)_{ee}^*$.
For future use, we also list
\eqali{
  \label{M:mutau}
(-++):&& 26\,\text{meV}&\lesssim (M_\nu)_{\mu\tau}\lesssim 28\,\text{meV}\,,\cr
(+-+):&& 20.5\,\text{meV}&\lesssim (M_\nu)_{\mu\tau}\lesssim 23\,\text{meV}\,.
}

\begin{figure}[h]
\centering
\includegraphics[scale=0.386,angle=0]{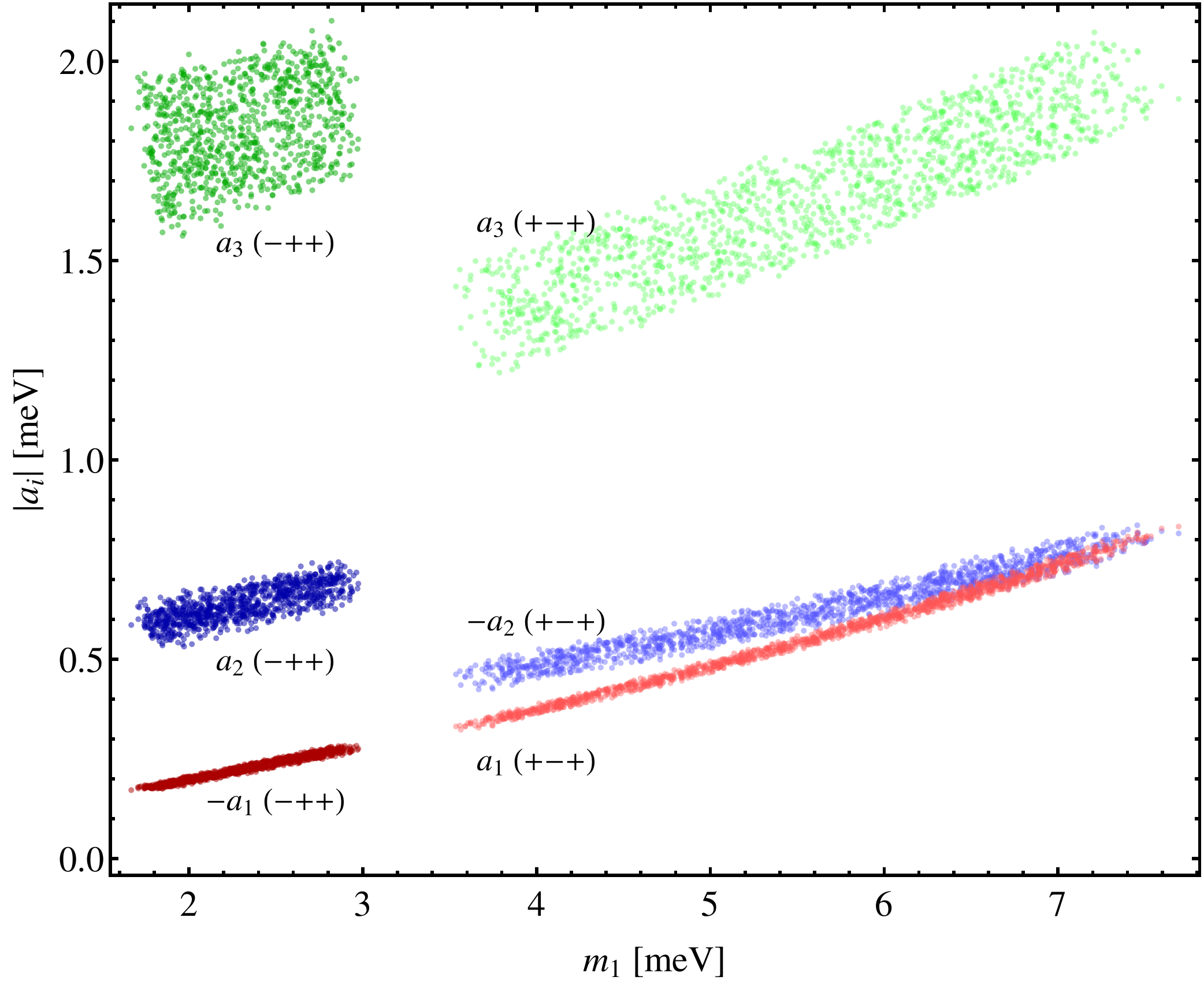}
\includegraphics[scale=0.384,angle=0]{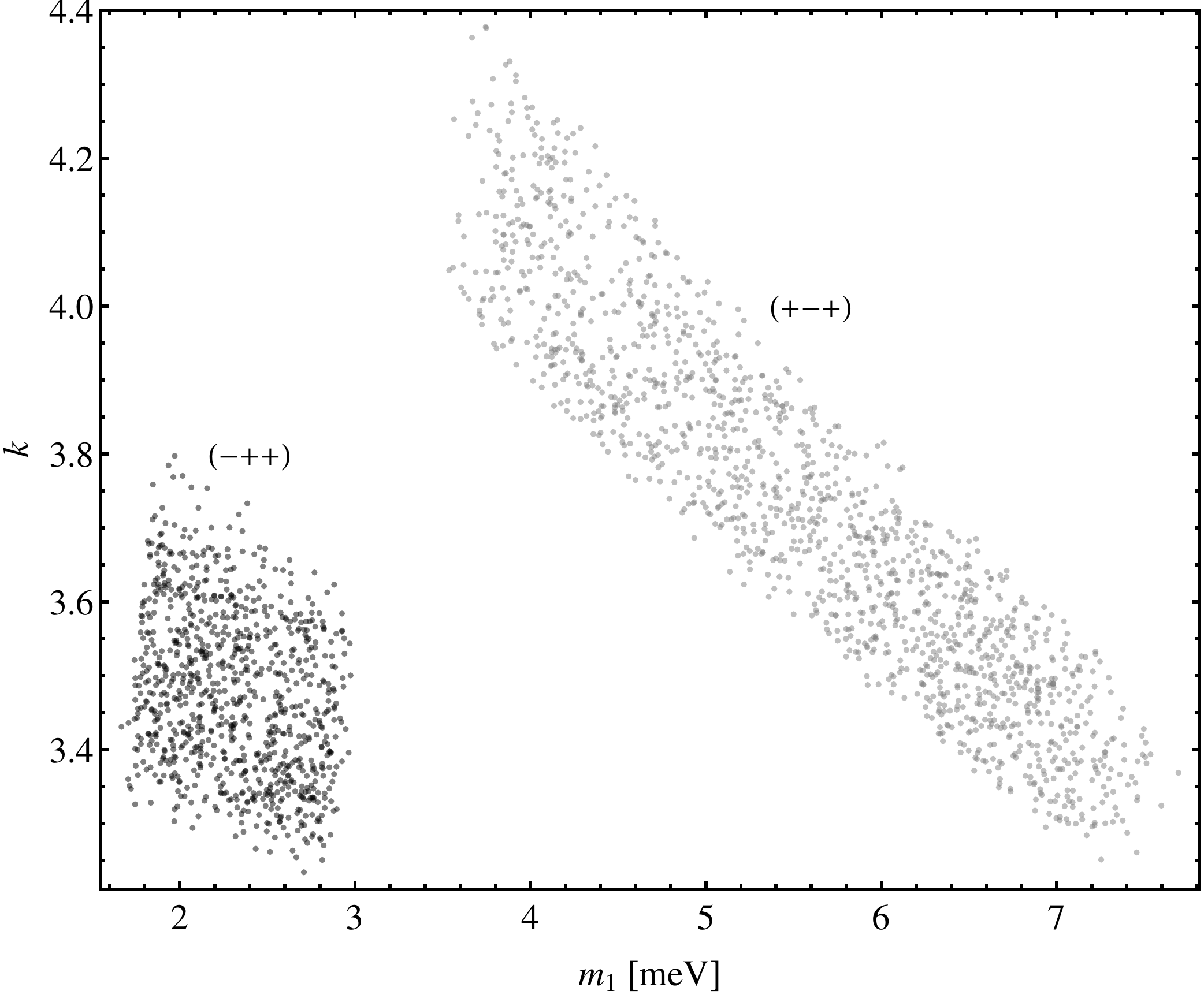}
\caption{
  \textbf{Left}: Scatter plot of $|a_i|$ as a function of the lightest mass $m_1$ 
(NH)
for the two possible CP parities for the light $\nu_{iL}$: $(-++)$ 
(darker colors) and $(+-+)$ (lighter colors).
Other solutions are related by permutations of $a_i$, cf. \eqref{prop:Mnu}. 
The depicted ordering of $a_i$ leads to $\delta_\cp=-\pi/2$.
\textbf{Right}: $k$ as a function of $m_1$; black dots and gray dots denote the 
cases $(-++)$ and $(+-+)$ respectively.
}
\label{fig:ai-k}
\end{figure}

The parameter distribution for the two sets of solutions is shown in
Fig.\,\ref{fig:ai-k} for $|a_i|$ as functions of $m_1$ (left), and $k$ as a 
function of $m_1$ (right). The values $\theta_{23}=\pi/4$ and $\dcp=\pm \pi/2$ are 
fixed from symmetry and we only consider values for $\theta_{12},\theta_{13},\Delta 
m^2_{21}$ and $\Delta m^2_{23}$ within 3-$\sigma$ of the global fit in 
Ref.\,\cite{GG:fit} by varying $a_i$ and $k$ independently.
Approximate values are obtained from the procedure below.

\begin{figure}[h]
\centering
\includegraphics[scale=0.386,angle=0]{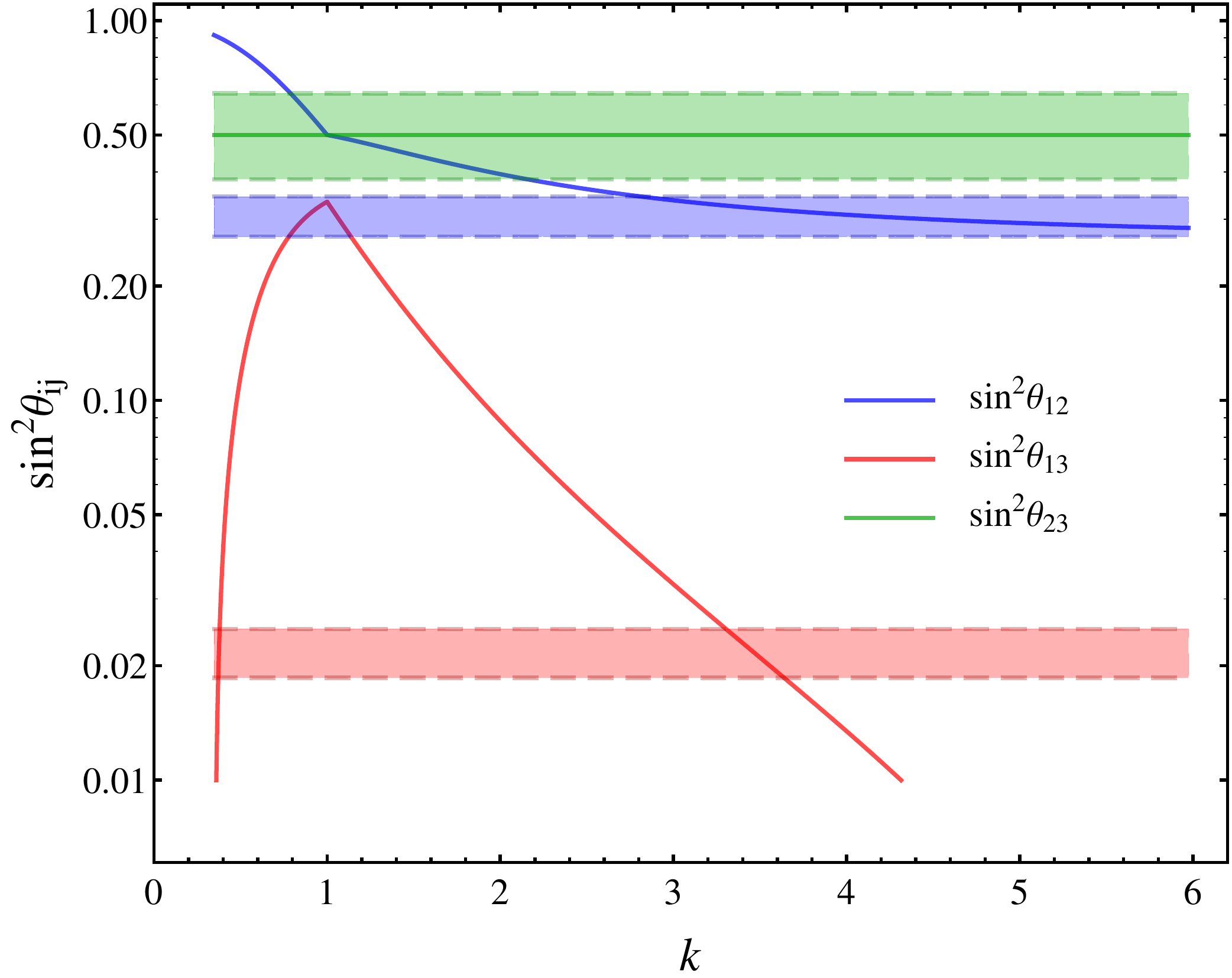}
\caption{
$\sin^2\theta_{ij}$ as a function of $k$ for $m_1=2.5\rm\,meV$. 
The colored bands corresponds to the allowed ranges of $\sin^2\theta_{ij}$.
We use the same parameters as Fig.\,\ref{fig:miai} and the procedure is explained 
in Sec.\,\ref{sec:sols}.
}
\label{fig:s2k}
\end{figure}

We use the following procedure to exclude solutions and 
search for approximate solutions:
\begin{enumerate}
\item For each lightest mass $m_0$, we find $\ta_i$ through Eq.\,\eqref{ai->mi:2} 
(or Eq.\,\eqref{cubic} with \eqref{c-tilde}) for a given $k$, restricted to the 
range specified by Fig.\,\ref{fig:kcrit}. We keep $\Delta m^2_{12}$ and $\Delta 
m^2_{23}$ fixed to their best-fit values of \cite{GG:fit}. An illustration of this 
procedure is given in Fig.\,\ref{fig:miai}.

\item Then, we diagonalize \eqref{Mnu:a4cp} to extract the mixing matrix $U=U_{\rm 
PMNS}$. We adopt the ordering of eigenvectors to satisfy
\eq{
  \label{order}
|U_{e1}|>|U_{e2}|>|U_{e3}|\,.
}
The ordering of $m_i$ follows.
This means that our mass eigenstates $\nu_1,\nu_2,\nu_3$ are in the order of 
decreasing contribution to $\nu_e$ ($\nu_1$ contributes the most and so on) and not 
in a specific mass ordering.
This definition explains the color flipping in Fig.\,\ref{fig:miai} for $k<1$.

\item At last, we check if the mass ordering is correct and if the mixing angles 
fall inside the 3-$\sigma$ ranges.
An illustration of this step is shown in Fig.\,\ref{fig:s2k}.
\end{enumerate}

One remark on this procedure is in order: to correctly fit the 
oscillation parameters we need that (i) the mixing angles are correct and (ii) the 
mass ordering is correct.
The condition (ii) arises because mass eigenstates $\nu_i$ are defined by 
\eqref{order} and mass orderings that do not correspond to NH or IH are excluded.
For example, we can read from Fig.\,\ref{fig:s2k} that the correct values for both
$s_{12}^2$ and $s^2_{13}$ are only achieved for $k\approx 3.5$, as can also be 
confirmed in Fig.\,\ref{fig:ai-k}.
Correct values for $s^2_{13}$ can also be obtained for $k\approx 0.4$ but 
$s^2_{12}$ as well as the mass ordering in Fig.\,\ref{fig:miai} is not correct: 
for $k<1$, $\nu_e$ has a greater contribution from the heaviest state ($\nu_2$ 
in red) than the second heaviest state ($\nu_3$ in green).

\section{Extended seesaw model}
\label{sec:ESS}

Here we present a low-scale seesaw model where the light neutrino 
mass matrix has the form \eqref{Mnu:a4cp}.
The model will retain the successful predictions of 
$\Umutau\times\ZZ_2^\cp$\,\cite{cp.mutau} for the low-energy neutrino 
observables but additional predictions arise due to the more constrained nature of 
the group $A_4$.
Two sets of heavy neutrinos -- one at the GeV-scale and another at the electroweak 
scale -- arise naturally due to the extended seesaw mechanism (ESS)\,\cite{ess}.
The combination of lepton flavor numbers $\mathtt{L}_\mu-\mathtt{L}_\tau$ will be 
approximately conserved in the model.

The flavor symmetry of the model will be $A_4\times\ZZ_2^\cp$ ($\cp_2$), explained 
in Sec.\,\ref{sec:a4}\,%
\footnote{Note that the combined group is a direct product because both factors 
commute\,\cite{cp.mutau}.}.
The SM lepton fields are, however, all singlets of $A_4$ and only feel the $\ZZ_3$ 
subgroup\,%
\footnote{This contrasts with most of the $A_4$ models for leptons where 
at least the lepton doublets form triplets\,\cite{review}
}, 
thus entirely avoiding the need of any vev alignment in this sector.
An auxiliary $\ZZ_4^D$ will also be necessary in the neutrino sector.
The SM lepton fields are assigned to $L_i\sim l_i\sim (\bs{1},\bs{1}',\bs{1}'')$ 
while the Higgs doublet $H$ is invariant; $L_i=(\nu_{iL},l_{iL})^\tp$ are lepton 
doublets while $l_i\equiv l_{iR}$ are the charged lepton singlets.
Thus $\cp_2$ in \eqref{cp2:charged} can be identified with 
$\cpmutau$\,\cite{cp.mutau}.
There are also two sets of SM singlets (right-handed neutrinos) $N_i\equiv N_{iR}$ 
and $S_i\equiv S_{iR}$, $i=1,2,3$, assigned to $(\bs{1},\bs{1}',\bs{1}'')$ and 
$\bs{3}$ respectively.
Hence, only the neutrino sector feels the full $A_4$ group through 
$S_{iR}$.
We also need complex flavons $\eta\sim \bs{3}$ and $\varphi_1\sim \bs{1}'$, and a 
real $\varphi_0\sim \bs{1}$.
The full assignment can be seen in table \ref{table:charges}.
Additional fields necessary to break $\cpmutau$ in the charged lepton sector are 
not shown since they can just be adapted from \cite{cp.mutau}.
\begin{table}
\eq{\nonumber
\begin{array}{|c|cccccccc|}
\hline
   & L_i & l_i & H & N_i & S_i & \eta & \varphi_0 & \varphi_1 \\
\hline
 A_4 & (\bs{1},\bs{1}',\bs{1}'') & (\bs{1},\bs{1}',\bs{1}'') & \bs{1} & 
(\bs{1},\bs{1}',\bs{1}'')& \bs{3} & \bs{3} & \bs{1} & \bs{1}' \\
 \ZZ_4^D & 1 & 1 & 1 & 1 & i & -i & -1 & -1 \\
\hline
\end{array}
}
\caption{
\label{table:charges}
Representation assignments for the fields.
}
\end{table}

The charged lepton sector at the electroweak scale will effectively be the SM 
one\,%
\footnote{%
For simplicity we are considering the UV completion by heavy leptons but the 
multi-Higgs version can be equally considered with the difference that the Higgs 
that couples to the $\mu-\tau$ flavors is distinct\,\cite{cp.mutau}.}
\eq{
\label{lag:l:eff}
-\lag^l_{\rm eff}=y_1\bar{L}_1Hl_1
+y_2\bar{L}_2 Hl_2 +y_3\bar{L}_3 Hl_3
+h.c.
\,,
}
where the $\ZZ_3$ subgroup is unbroken but $\cpmutau$ is broken at a higher scale 
by a CP-odd scalar\,\cite{cp.mutau} so that the correct splitting for 
$y_\mu=y_2$ and $y_\tau=y_3$ is generated ($y_e=y_1$).

The neutrino sector at the high scale is given by
\eqali{
  \label{lag:ESS}
-\lag_\nu&=f_1\bar{N}_{1}\tH^\dag L_1
  +f_2\bar{N}_{2}\tH^\dag L_2+f_3\bar{N}_{3}\tH^\dag L_3
\cr  &~~
+ f_1'(\bS\eta)_{1}N_1^c
+ f_2'(\bS\eta)_{1'}N_2^c
+ f_3'(\bS\eta)_{1''}N_3^c
\cr  &~~
+ \ums{2}M_{11}\bN_1N_1^c + M_{23}\bN_2N_3^c
\cr  &~~
+ \ums{2}k_0\varphi_0(\bS S^c)_{1}
+ \ums{2}k_1\varphi_1(\bS S^c)_{1''}
+ \ums{2}k_1^*\varphi_1^*(\bS S^c)_{1'}
\cr  &~~
+h.c.,
}
where we have defined singlet combinations of two triplets of $A_4$ as
\eqali{
\label{singlets}
(xy)_{1}&\equiv x_1y_1+x_2y_2 +x_3y_3\,,\cr
(xy)_{1'}&\equiv x_1y_1+x_2y_2\om^2 +x_3y_3\om\,,\cr
(xy)_{1''}&\equiv x_1y_1+x_2y_2\om +x_3y_3\om^2\,.
}
Note that $\cpmutau$ acts as
\eqali{
\cpmutau:& \quad
L_1\to L_1^{cp},\quad
L_2\to L_3^{cp},\quad
L_3\to L_2^{cp},
\cr
&\quad H \to H^*,\quad
S_i\to S_i^{cp},\quad
\eta_i\to \eta_i^*,
\cr
&\quad 
\varphi_0\to \varphi_0,\quad
\varphi_1\to \varphi_1\,,
}
and $l_i$ and $N_i$ transform like $L_i$ and $\psi^{cp}$ denotes the usual CP 
conjugate of the chiral fermion $\psi$.
Therefore, $f_1,f_1',M_{11},M_{23},k_0$ are real and $f_3=f_2^*$, ${f_3'}^*=f_2'$ 
due to $\cpmutau$.
The parameters $f_{2,3},f'_{2,3}$  can be further chosen real and positive by 
rephasing $L_{2,3}$ and $N_{2,3}$.

The mass matrix for $(\nu_{iL},N_i^c,S_i^c)$ after EWSB will be
\eq{
    \label{MM}
\bbM=\mtrx{\bs{0}& M_D^\tp\cr M_D & M_R}
=
\mtrx{\bs{0}& m_D^\tp&\bs{0}\cr 
 m_D& M_N&\Lambda^\tp\cr
\bs{0}&\Lambda &\mu }\,,
}
where
\eqali{
\label{m.matrix}
 m_D&=\diag(\md{i})=\frac{v}{\sqrt{2}}\diag(f_1,f_2,f_3)\,,\cr
\Lambda&=\diag(u_1,u_2,u_3)\sqrt{3}U_\om^*\diag(f_1',f_2',f_3')\,,\cr
M_N&=\mtrx{M_{11}&&\cr&&M_{23}\cr &M_{23}&}\,,\cr
\mu&=\diag(\mu_1,\mu_2,\mu_3)\,,
}
where
\eq{
U_\om\equiv \frac{1}{\sqrt{3}}\mtrx{1&1&1\cr 1&\om&\om^2\cr 1&\om^2 &\om}\,.
}
In this model, we are considering that $\varphi_{0,1}$ acquire very 
small vevs which lead to the real Majorana masses $\mu_i$ for $S_i$ and also 
\eq{
\aver{\eta_i}=u_i\,,\quad \text{all real.}
}
We justify the hierarchy of vevs in appendix \ref{ap:pot}.

Considering that $ M_N$ is composed of bare masses, the ESS limit is naturally 
achieved\,\cite{ess}: $ M_N\gg \{\Lambda, m_D\}\gg \mu$ and also $\mu\ll 
\Lambda^2/M_N$.
We can see that there are two sources of lepton number violation (LNV) in 
\eqref{lag:ESS}\,%
\footnote{If $\mathtt{L}(N_i)=\mathtt{L}(L_i)=-\mathtt{L}(S_i)=1$.}%
: (a) large scales $M_N$ and (b) low-scales $\mu_i\sim 
\aver{\varphi_{0,1}}$.

At tree level and leading order we obtain
\eqali{
    \label{ess:m}
\nu_i:&& M_\nu&=  m_D^\tp\Lambda^{-1}\mu (\Lambda^{\tp})^{-1} m_D\,,
\cr
S_i^c:&& M_S&=-\Lambda M_N^{-1}\Lambda^\tp\,,
\cr
N_i^c:&& M_N& \,,
}
with light-heavy mixing
\eqali{
\label{nu-SN}
\theta_{\nu S}^*&= m_D^\tp \Lambda^{-1}\,,
\cr
\theta_{\nu N}^*&= m_D^\tp M_N^{-1}\,.
}
Additional mixings can be seen in appendix \ref{ap:block}.
We can see that the small LNV scale $\mu$ only enters $M_\nu$ while the large LNV 
scale $M_N$ contributes only to heavier masses.
Given that the mass matrix for the heavier states $N_i$ are approximately 
unchanged, we can define
\eq{
M_{N_1}\equiv M_{11},\quad
M_{N_2}=M_{N_3}\equiv M_{23},
}
assuming positive quantities.
The leading correction can be seen in appendix \ref{ap:block}.

Explicitly, the light neutrino mass matrix is
\eq{
    \label{mnu:tree}
M_\nu=\ums{3}\diag(\md{i}/f_i')U_\om \diag(\mu_i/u_i^2)U_\om \diag(\md{i}/f_i')\,,
}
which has the desired form \eqref{Mnu:a4cp} with
\eq{
    \label{ai:model}
a_i=\ums{9}\mu_i\frac{\md1^2}{u_i^2f_1'^2}\,,\quad
k=\frac{|\md2 f_1'|}{|\md1 f_2'|}\,.
}
We have used the shorthand $\md1 \equiv f_1v/\sqrt{2}$; cf. \eqref{m.matrix}.
The fitting of the light neutrino parameters in Fig.\,\ref{fig:ai-k} implies 
\eqali{
\label{fit:ranges}
|a_i|&\approx 0.2 \text{ -- } 2.1\,\text{meV}\,,
\cr
k&\approx 3.2 \text{ -- }4.4\,.
}
Also, the sign change of one of the $a_i$ needs to be generated by $\mu_i$ and 
not by $u_i^2$ which is always positive.

Although the heavier states $N_i$ are frequently chosen to lie above the TeV 
scale\,\cite{vissani,ess}, in our case (i) a negligible one-loop contribution for 
light neutrino masses, (ii) validity of the ESS approximation and (iii) BBN 
constraints will essentially restrict $M_N$ to the electroweak scale; see 
Sec.\,\ref{sec:one-loop} and \eqref{bench} for a benchmark point.

\section{Approximate $\Umutau$ limit}
\label{sec:Umutau}

We consider first the limit where $\ZZ_3$ of $A_4$ is only broken by the small 
quantities in $\mu$. 
This means that below the scale of $\aver{\eta}$, $\ZZ_3$ is only broken by light 
neutrino masses.
This approximate $\ZZ_3$ symmetry corresponds to the \textit{lepton flavor 
triality} (LFT)\,\cite{lep.triality} where lepton fields carry the discrete charges
\eq{
\label{triality}
\text{LFT}:\quad L_i\sim l_i\sim N_i \sim S_i'^c \sim (1,\om,\om^2)\,;
}
$S_i'$ is related to $S_i$ by change of basis $S_i^c=(U_\om)_{ij} S_j'^c$.

The heavy vevs of $\eta$ conserve LFT when
\eq{
  \label{111}
\aver{\eta_i}=u_i\approx u_0(1,1,1)\,.
}
This feature is justified in appendix \ref{ap:pot}.
In this case, after $\eta_i\to \aver{\eta_i}$ and in the limit $k_1\to 0$, the 
Lagrangian \eqref{lag:ESS} is in fact invariant by the continuous version of 
\eqref{triality} with charges\,\cite{cp.mutau}
\eq{
\label{mu-tau}
\Umutau:\quad L_i\sim l_i\sim N_i \sim S_i'^c \sim (0,1,-1)\,.
}
It corresponds to the combination $\mathtt{L}_\mu-\mathtt{L}_\tau$ of family lepton 
numbers.
The approximate conservation of $\Umutau$ will lead to a number of consequences.

In this limit the mass matrix \eqref{ess:m} and mixing \eqref{nu-SN} of the 
heavy neutrinos $S_i$ yield
\eqali{
\label{MS:1}
 M_S&=-\ums{3}
\mtrx{\Ms{1}+2\Ms{2} & \Ms{1}-\Ms{2} & \Ms{1}-\Ms{2} \cr 
\star & \Ms{1}+2\Ms{2} & \Ms1-\Ms2 \cr
\star & \star & \Ms1+2\Ms2}\,,
\cr
\theta_{\nu 
S}^*&=\frac{\md1 }{\sqrt{3}f_1'u_0}\mtrx{1&&\cr&k&\cr&&k}
U_\om
\,,
}
where the masses read\,\footnote{%
We keep using the same name $S_i$ for the heavy neutrino fields although they have 
a small component of $\nu_{iL}^c$ and $N_{iR}$.
}
\eq{
\label{masses:S}
\Ms1\equiv\frac{(\sqrt{3}u_0f_1')^2}{M_{N_1}}\,,\quad
\Ms{2,3}\equiv\frac{(\sqrt{3}u_0|f_2'|)^2}{M_{N_2}}\,.
}
These relations allows us to trade $f_1'u_0$ and $f_2'u_0=f_3'u_0$ for physical 
masses:
\eq{
\label{trade:LL}
|\sqrt{3}f_1'u_0|=\sqrt{\Ms1 M_{N_1}}\,,\quad
|\sqrt{3}f_2'u_0|=\sqrt{\Ms2 M_{N_2}}\,.
}

The mass matrix $M_S$ is invariant by cyclic permutations and then $(1,1,1)$ is an 
eigenvector.
We can diagonalize it by
\eq{
    \label{VS}
V_S^*=U_\om^*(-i) U_{23}^*
\,.
}
giving
\eq{
V_S^\tp M_S V_S= \diag(\Ms1,\Ms2,\Ms3)\,.
}
The matrix $U_{23}$ was defined in \eqref{U23}.
Therefore, $S_1$ is a Majorana fermion of $\Umutau$ charge 0 and $S_{2,3}$ are 
degenerate Majorana fermions that form a (pseudo-)Dirac pair of fields with charge 
$\pm 1$. The latter implies that LNV effects induced by $S_{2,3}$ exchange will 
vanish in this limit.

The active-sterile $\nu$-$S$ mixing reduces to
\eq{
  \label{nu-S}
(\theta_{\nu S}V_S)^*=
(-i)\frac{\md1}{\sqrt{3}u_0f_1'}\diag(1,k,k)
\left(
\begin{array}{ccc}
 1 & 0 & 0 \\
 0 & \frac{1}{\sqrt{2}} & -\frac{i}{\sqrt{2}} \\
 0 & \frac{1}{\sqrt{2}} & \frac{i}{\sqrt{2}} \\
\end{array}
\right)
\,.
}
It is important to note that in this approximation
\eq{
  \label{theta.ei}
(\theta_{\nu S}V_S)_{ei}=0\,,\text{ for $i=2,3$},
}
and the electron flavor is only coupled to $S_1$.

\section{One-loop contributions and BBN constraints}
\label{sec:one-loop}

Now we should compute the one-loop contributions to light neutrino masses.
When the lightest heavy RHN mass lies below 100 MeV, the one-loop contributions 
to $\nuless$ can be sizable\,\cite{lopez-pavon}, although such a sterile neutrinos 
are severely constrained by cosmological data\,\cite{cosmo}.
Heavy neutrinos with electroweak-scale masses can still induce sizable 
contributions\,\cite{ibarra.petcov,vissani} and the dominant (and finite) one comes 
from light neutrino self-energies with Higgs or $Z$ 
exchange\,\cite{1-loop,aristizabal,petcov.15}.

We can write the self-energy contribution as
\eq{
\label{1-loop:type-I}
M_\nu^{\onel}=\frac{1}{(4\pi v)^2}M_D^\tp \Big(
M_R^{-1}F(M_RM_R^\dag)+F(M_RM_R^\dag)M_R^{-1}
\Big)
M_D\,,
}
where the loop function $F(x)$ is given by
\eq{
F(x)\equiv \frac{x}{2}\left[
3\,\frac{\ln(x/M_Z^2)}{x/M_Z^2-1}
+\frac{\ln(x/M_h^2)}{x/M_h^2-1}
\right]\,,
}
with $M_Z$ and $M_h$ being the $Z$ and Higgs boson masses, respectively;
$v=246\,\unit{GeV}$ is the electroweak scale.
This contribution should be added to the tree-level contribution \eqref{mnu:tree} 
coming from the ESS mechanism.
We should note that heavy neutrino masses $M_R$ at the electroweak scale leads to a
contribution \eqref{1-loop:type-I} functionally similar to the tree-level 
contribution $M_D^2/M_R$, but smaller only by the loop factor 
$1/16\pi^2$\,\cite{1-loop} [notice $F\big((100\,\text{GeV})^2\big)/v^2\approx 1.5$].
Therefore, the one-loop contribution in the ESS mechanism can possibly be large 
since the cancellation that occurs in the tree-level mass matrix is not 
expected to carry over to the one-loop contribution.

We can adapt the one-loop contribution for generic type-I 
seesaw \eqref{1-loop:type-I} to the extended seesaw with mass matrix \eqref{MM} as
\eq{
\label{1-loop:ess:1}
M_\nu^{\onel}=\frac{1}{(4\pi v)^2}m_D^\tp
\bigg\{
    M_N^{-1}\Lambda^\tp V_S\,\hM_S^{-1} 2F(\hM_S^2) V_S^\tp \Lambda M_N^{-1}
+V_N\,\hM_N^{-1} 2F(\hM_N^2) V_N^\tp
\bigg\}
m_D\,.
}
We have first block diagonalized $M_R$ (see appendix \ref{ap:block}) and then 
used the basis where $M_S$ and $M_N$ is diagonal ($\hM_S$ and $\hM_N$).
It is also possible to write the expression in terms of the light-heavy mixing 
angles as
\eq{
\label{1-loop:ess:2}
M_\nu^{\onel}=\frac{1}{(4\pi v)^2}\bigg\{(\theta_{\nu S} V_S)^*
    \hM_S\,2F(\hM_S^2) (\theta_{\nu S} V_S)^\dag
+(\theta_{\nu N}V_N)^*\hM_N\,2F(\hM_N^2) (\theta_{\nu N}V_N)^\dag
\bigg\}\,.
}
We can see that generically the contribution from the heavier states $N_i$ 
\textit{dominates} over the contribution from $S_i$ because the smaller mixing angle
$\theta_{\nu N}^2/\theta_{\nu S}^2\sim \Lambda^2/M_N^2$ is compensated by 
$M_N/M_S\sim M_N^2/\Lambda^2$ and $F(x)$ grows with $x$.

\begin{figure}[h]
\centering
\includegraphics[scale=0.41,angle=0]{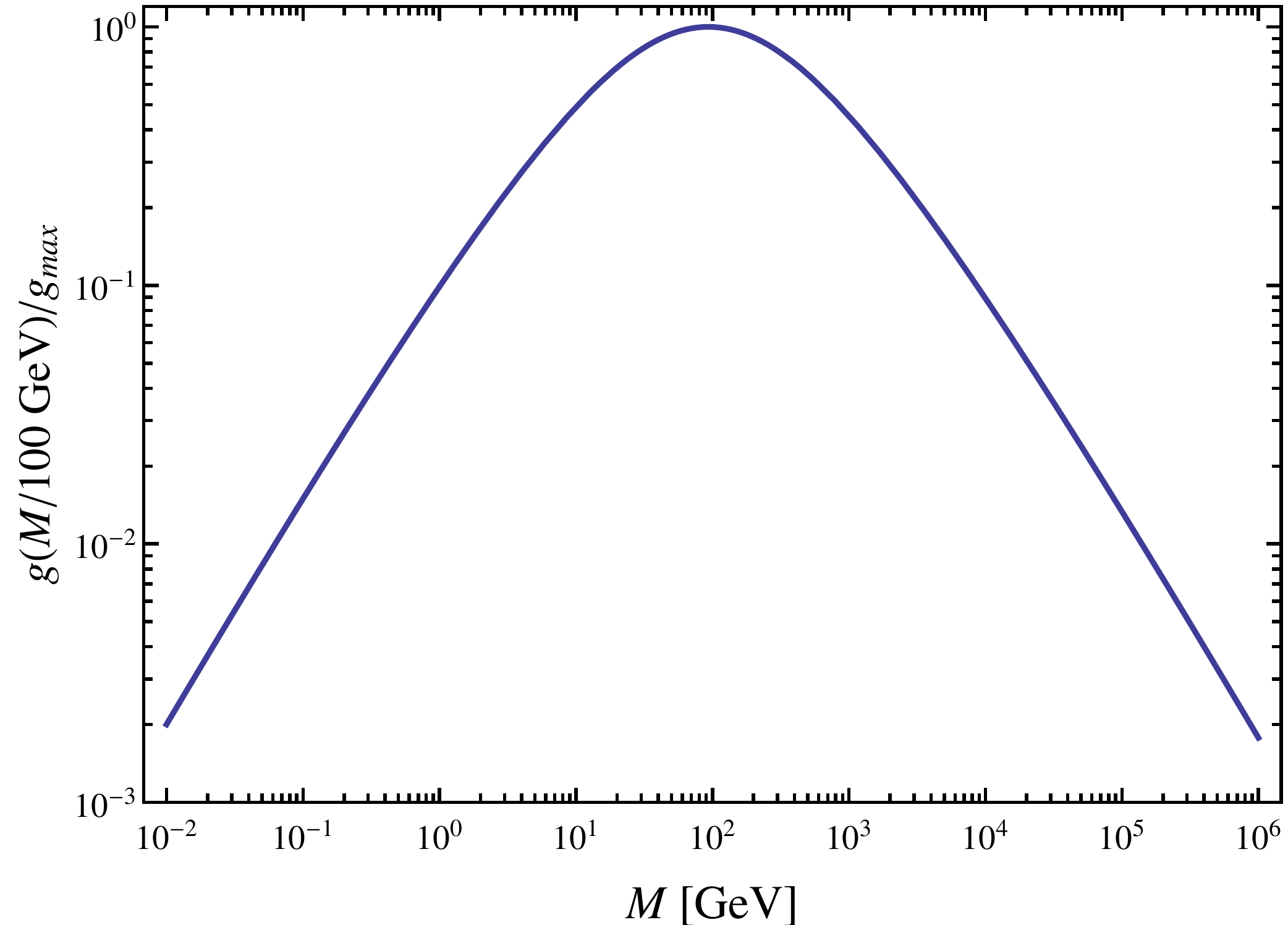}
\caption{
Plot of the function in \eqref{def:g} where $g_{\max}=3.63547$ and the maximum 
occurs at $M\approx 93.3\,\unit{GeV}$.
}
\label{fig:g}
\end{figure}

For our purposes, it is useful to define the adimensional function $g(x)$ as
\eq{
    \label{def:g}
g(M_i/100\,\text{GeV})\equiv 
\frac{2F(M_i^2)}{M_i\times 100\,\text{GeV}}\,.
}
A slightly different definition can be seen in \cite{rad.iss}.
This function peaks at the electroweak scale $M_i\approx 93.3\,\text{GeV}$ with 
maximum $3.64$ and decreases away from the peak with rate slower than $M_i^{-1}$ 
for $M_i\gtrsim 100\,\text{GeV}$; see behaviour in Fig.\,\ref{fig:g}.
This function allows us to rewrite \eqref{1-loop:ess:1} as
\eq{
\label{1-loop:ess}
M_\nu^{\onel}=\frac{100\,\text{GeV}}{(4\pi v)^2}m_D^\tp
\bigg\{
    M_N^{-1}\Lambda^\tp V_S\,g(\hX_S) V_S^\tp \Lambda M_N^{-1}
+V_N\,g(\hX_N) V_N^\tp
\bigg\}
m_D\,.
}
We have used the shorthand $\hX_S\equiv \diag(M_{S_i})/100\,\text{GeV}$ and 
similarly for $\hX_N$.

Computing \eqref{1-loop:ess} in our model in the $\Umutau$ symmetric limit, we 
obtain the texture
\eq{
M_\nu^{\onel}=\mtrx{\star &0 &0\cr 0&0&\star \cr 0&\star&0 }\,,
}
whose nonzero entries correspond to $\mathtt{L}_\mu-\mathtt{L}_\tau=0$.
Explicitly, 
\eqali{
    \label{1-l:model}
(M_\nu^{\onel})_{ee}&\approx 10\,\unit{keV}\times
\frac{\md1^2}{\unit{GeV}^2}
\Big[-\frac{g(x_1)}{M_{N_1}/M_{S_1}}+g\Big(\frac{M_{N_1}}{M_{S_1}} x_1\Big)\Big]\,,
\cr
(M_\nu^{\onel})_{\mu\tau}&\approx 10\,\unit{keV}\times
\frac{\md2^2}{\unit{GeV}^2}
\Big[-\frac{g(x_2)}{M_{N_2}/M_{S_2}}+g\Big(\frac{M_{N_2}}{M_{S_2}} x_2\Big)\Big]\,,
}
where $x_i\equiv\Ms{i}/100\,\rm GeV$.
We have used Eqs.\,\eqref{m.matrix}, \eqref{VS} and $V_N=U_{23}$.
We note that indeed the one-loop contribution can lead to an unacceptably large 
contribution. For example, for $m_D\sim 1\,\unit{GeV},M_N\sim 
10\,\unit{TeV},M_S\sim 100\,\unit{GeV}$, the one-loop contribution leads to a few 
keV.
From Fig.\,\ref{fig:g} we also see that to lower the
contributions from \eqref{1-l:model} to acceptable values by increasing $M_N$ 
requires very large values of the order of $10^7\,\text{GeV}$.
Therefore, to have TeV-scale (or lower) right-handed neutrinos, we need to lower 
the scale of $m_D$ or arrange some cancellation between either the various one-loop 
contributions or between the tree and one-loop ones\,\cite{petcov.15}.
We consider this possibility unappealing and do not pursue it any further.

\begin{figure}[h]
\centering
\includegraphics[scale=0.38,angle=0]{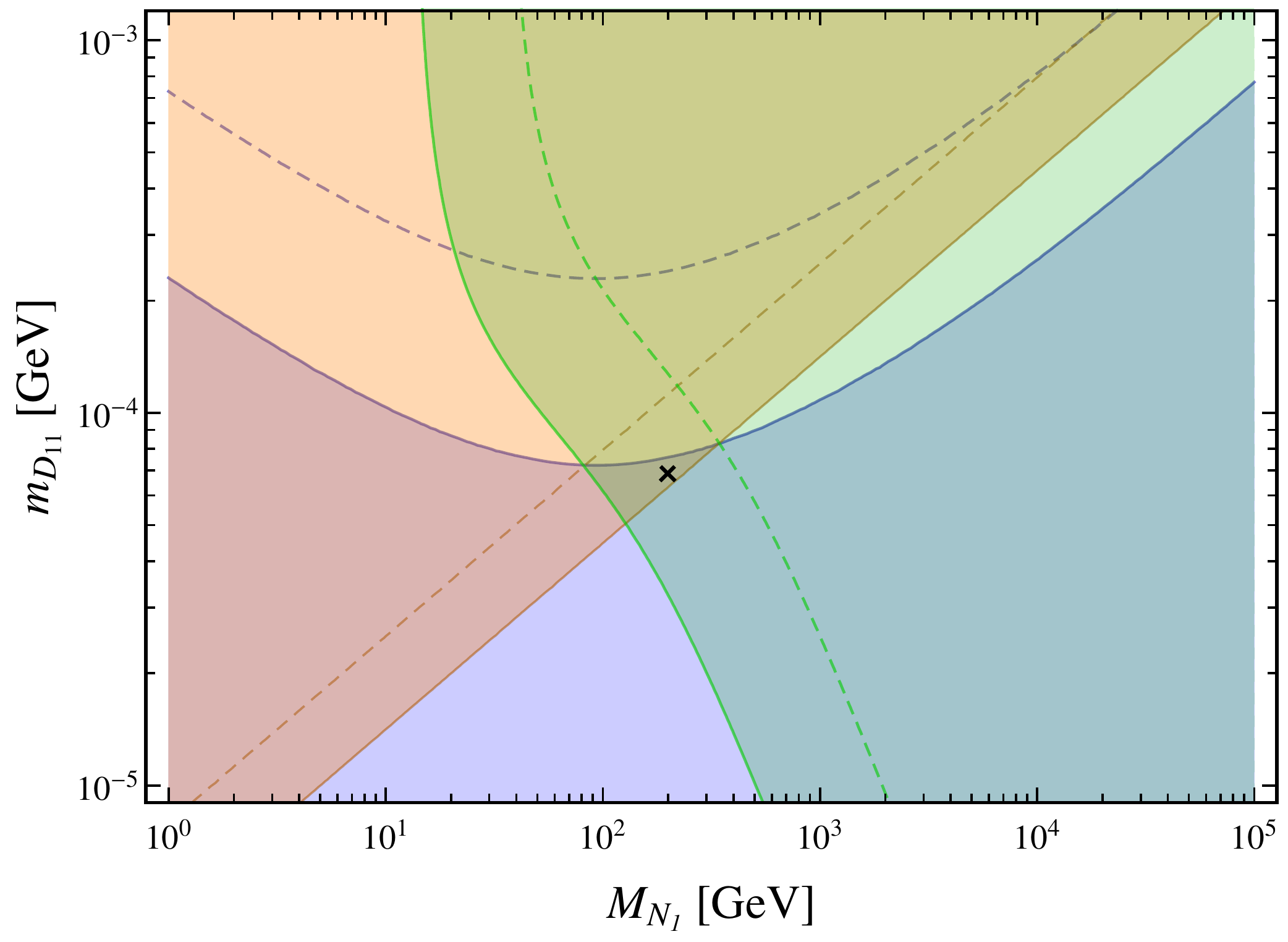}
\includegraphics[scale=0.38,angle=0]{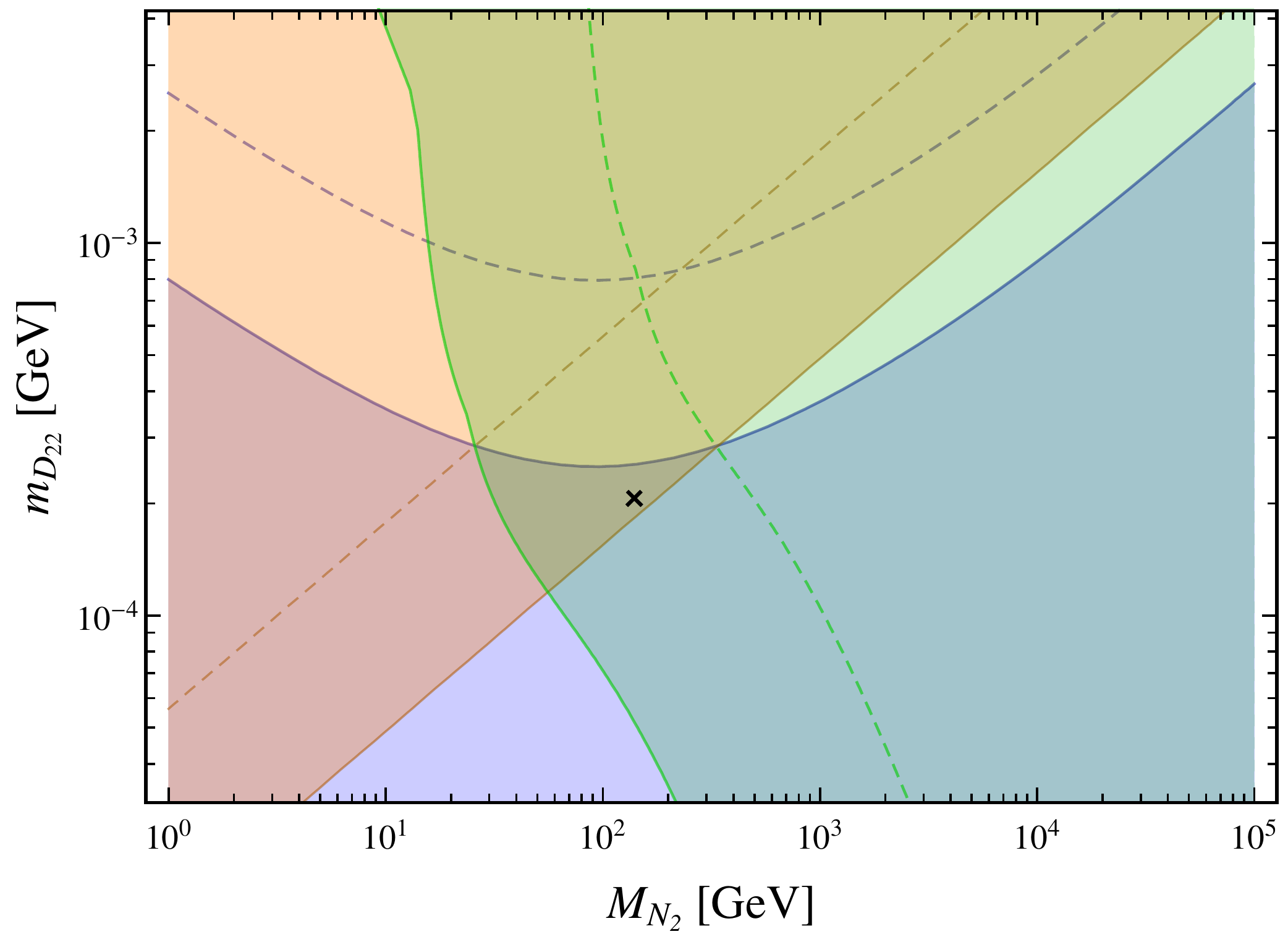}
\caption{
The blue regions satisfy 
$|(M_\nu^{\onel})_{ee}|/|(M_\nu^{\rm tree})_{ee}|\le 0.1$ (left) or
$|(M_\nu^{\onel})_{\mu\tau}|/|(M_\nu^{\rm tree})_{\mu\tau}|\le 0.1$ (right) 
with the reference values \eqref{ref.value}.
The blue dashed curves obey unit ratios.
The points inside the orange regions are the ones necessary to fit the 
$ee$ (left) or $\mu\tau$ (right) tree-level entries of the light neutrino mass 
matrix through \eqref{ai:mag:2} or \eqref{ai:mag:3} restricted to \eqref{Rmu} 
and \eqref{ref.value}.
The orange dashed curves correspond to the subset of points for $R_{\mu1}=0.03$ 
(left) and $R_{\mu2}=0.076$ (right).
The green regions cover the points where $\tau_{S_1}\le 0.1$ (left) or 
$\tau_{S_2}\le 0.1$ (right) for $R_{N_1}=R_{N_2}=100$.
The green dashed curves yields the lifetime of 0.1\,s but with $R_{N_1}=270$ 
(left)
or $R_{N_2}=400$ (right).
The crosses mark the benchmark points in \eqref{bench}.
See text for details.
}
\label{fig:MDMN}
\end{figure}

In order to preserve our predictions of Sec.\,\ref{sec:mass} we confine ourselves 
to 
the case where the loop-induced contributions \eqref{1-l:model} are negligible 
compared to the tree level ones in \eqref{Mnu:a4cp}.
To visualize the possible regions in parameter space, we show in 
Fig.\,\ref{fig:MDMN} the regions (blue) in the $M_{N_1}$-$\md1$ plane (left) and 
$M_{N_2}$-$\md2$ plane (right) where the one-loop contribution is at most 10\% of 
the tree-level contribution for the $ee$ (left) and $\mu\tau$ (right) entries.
For definiteness we fix the tree-level values to
\eqali{
\label{ref.value}
(M_{\nu}^{\rm tree})_{ee}&=2\,\unit{meV},\quad
R_{N_1}\equiv M_{N_1}/\Ms1=10^2\,,\cr
(M_{\nu}^{\rm tree})_{\mu\tau}&=24\,\unit{meV}\,,\quad
R_{N_2}\equiv M_{N_2}/\Ms2=10^2\,.
}
These values are in agreement with \eqref{nuless:0} and \eqref{M:mutau}.
We choose to plot the dependence on $M_{N_i}$ because the one-loop 
contributions depend dominantly on $M_{N_i}$ (rather than on the lighter $\Ms{i}$) 
in the ESS approximation.
For example, if we increase the ratios $R_{N_i}$, the blue regions shrinks down 
only slightly for large $M_{N_i}$.
For completeness, we also show the curves for unit ratio (dashed).

The next step is to ensure that the tree-level contribution themselves -- as they 
\textit{depend} on the model parameters as in \eqref{ai:model} -- 
lie in the necessary ranges of \eqref{nuless:0} and \eqref{M:mutau}
(also Fig.\,\ref{fig:ai-k}).
For that purpose, we rewrite the sum of all relations for $a_i$ in \eqref{ai:model} 
as
\eq{
\label{ai:mag:2}
\left(\frac{\md1}{10\text{keV}}\right)^2
\left(\frac{\text{100\,GeV}}{M_{N_1}}\right)
\left(\frac{\bar{\mu}_i}{M_{S_1}}\right)
=\frac{|(M^{\rm tree}_{\nu})_{ee}|}{\text{meV}}
\,,
}
where $\bar{\mu}_i\equiv\sum_i\mu_i/3$.
We have also used \eqref{trade:LL} to eliminate $f_1'u_0$.
An analogous relation is valid for the $\mu\tau$ entry:
\eq{
\label{ai:mag:3}
\left(\frac{M_{D_{22}}}{10\text{keV}}\right)^2
\left(\frac{\text{100\,GeV}}{M_{N_2}}\right)
\left(\frac{\bar{\mu}_i}{M_{S_2}}\right)
=\frac{|(M^{\rm tree}_{\nu})_{\mu\tau}|}{\text{meV}}
\,.
}
As $\bar{\mu}_i\ll \Ms1$ in order to satisfy the ESS approximation, we require 
\eq{
\label{Rmu}
R_{\mu1}\equiv\frac{\bar{\mu}_i}{\Ms1}\le 0.1
~\text{ and }~
R_{\mu2}\equiv \frac{\bar{\mu}_i}{\Ms2}\le 0.1
\,.
}
These conditions define allowed regions for $M_{N_1}$-$\md1$ and $M_{N_2}$-$\md2$
which are shown as orange regions in Fig.\,\ref{fig:MDMN}.
We also show in dashed orange curves the values where the above ratios assume the 
values $R_{\mu1}=0.03$ (left) and $R_{\mu2}=0.0076$ (right).
We use the same reference values in \eqref{ref.value}.

The conclusion is that the overlapping (allowed) regions impose upper bounds on 
the heavy RHN states:
\eq{
M_{N_1},M_{N_2}\lesssim 340\,\text{GeV}\,.
}
This constraint puts the RHN states $S_i$ at the GeV-scale.
We also note that had we allowed $M_\nu^{\onel}\sim M_\nu^{\rm tree}$, $M_{N_1}$ 
would be unbounded but restricted to a narrow band $M_{D_{11}}^2/M_{N_1}\sim 
10^{-11}\,\rm GeV$ for $M_{N_1}\gtrsim 1\,\rm TeV$.
A similar consideration applies to $M_{N_2}$.

As the last constraint, we note that $\md{i}$ cannot be pushed to arbitrarily low 
values because it necessarily makes the lighter BSM states $S_i$ very 
long-lived\,\footnote{%
We assume all the scalars to be heavier than $S_i$.
}.
In order to not spoil the successful prediction of Big Bang nucleosinthesis (BBN), 
we require that the lifetimes of all the BSM states do not exceed 0.1\,s.
It is enough to require that for the lighter $S_i$ states.
As their masses lie at the GeV-scale or lower, the main decay modes involve $W$ 
or $Z$ exchange through active-sterile mixing with decay into light neutrinos, 
electrons or pions\,\cite{shapo}; see appendix \ref{ap:decay} for more details.
The allowed regions are shown in green in Fig.\,\ref{fig:MDMN} where the border is 
determined by the fixed $N-S$ ratios of \eqref{ref.value}; the interior refers to 
$R_{N_1}>10^2$ (left) or $R_{N_2}>10^2$ (right) in accordance to the ESS 
approximation.
For completeness, we also show as dashed green curves the points where 
$\tau=0.1\,\text{s}$ and $R_{N_1}=270$ (left) or $R_{N_2}=400$ (right).

The combination of all the constraints discussed above, leads to the overlapping 
regions of Fig.\,\ref{fig:MDMN}.
The parameters are restricted to the values listed in Table \ref{table}.
The restriction means that points outside the overlapping region violate some 
constraint above for the reference values \eqref{ref.value}.%
\footnote{%
The actual green regions may lie slightly to the left for two reasons: 
(i) we only include the dominant decay modes for $S_i$ listed in appendix 
\ref{ap:decay} and (ii) the strict lifetime limit for successful BBN may be 
slightly relaxed depending on the details of the model at the BBN 
era\,\cite{nu:BBN}.
}
Points inside the overlapping regions need to be further checked for all the 
constraints as they depend on other parameters not shown in the figures.
Moreover, the parameters are not all independent as one ratio is fixed through 
\eqref{fit:ranges} and
\eq{
\frac{\md2}{\md1 }
\frac{M_{N_1}}{M_{N_2}}
\frac{\sqrt{R_{N_2}}}{\sqrt{R_{N_1}}}=k\,.
}
To use tree-level values different from \eqref{ref.value} but restricted to
\eqref{nuless:0} and \eqref{M:mutau}, we just need to reread Fig.\,\ref{fig:MDMN} 
with the vertical axis relabeled as
\eqali{
\md1&\to \md1\sqrt{\frac{2\,\text{meV}}{(M_{\nu})^{\rm tree}_{ee}} }\,,\cr
\md2&\to \md2\sqrt{\frac{24\,\text{meV}}{(M_{\nu})^{\rm tree}_{\mu\tau}} }\,.
}
This is possible because all the defining relations, Eqs.\,\eqref{1-l:model}
\eqref{ai:mag:2}, \eqref{ai:mag:3} and the active-sterile mixing $\theta_{\nu S}$ 
in the decay rates (ap.\,\ref{ap:decay}) depends on $\md1^2$ or $\md2^2$.
For the same reason, the blue and orange curves of the right figure of 
Fig.\,\ref{fig:MDMN} are identical to the ones on the left if we identify $\md2=
\md1\sqrt{24/2}$, where $\sqrt{24/2}$ is basically the factor $k$.
\begin{table}[h]
\eq{\nonumber
\begin{array}{|c|r|c|r|}
\hline
\md1 /10^{-5}\,\text{GeV} &  5\text{ -- }8 & 
    \md2/10^{-5}\,\text{GeV} & 12\text{ -- }28
   \\
M_{N_1}/\text{GeV} & 80\text{ -- }340 & 
    M_{N_2}/\text{GeV} & 25\text{ -- }340 \\
 M_{N_1}/M_{S_1} & 100\text{ -- }270  & 
    M_{N_2}/M_{S_2} & 100\text{ -- } 400 \\
\bar{\mu}_i/M_{S_1} & 0.03\text{ -- }0.1  & 
    \bar{\mu}_i/M_{S_2} & 0.0076\text{ -- }0.1  \\
\hline
\end{array}
}
\caption{\label{table}
Approximate parameter values extracted from Fig.\,\ref{fig:MDMN}.
}
\end{table}

\begin{figure}[h]
\centering
\includegraphics[scale=0.375,angle=0]{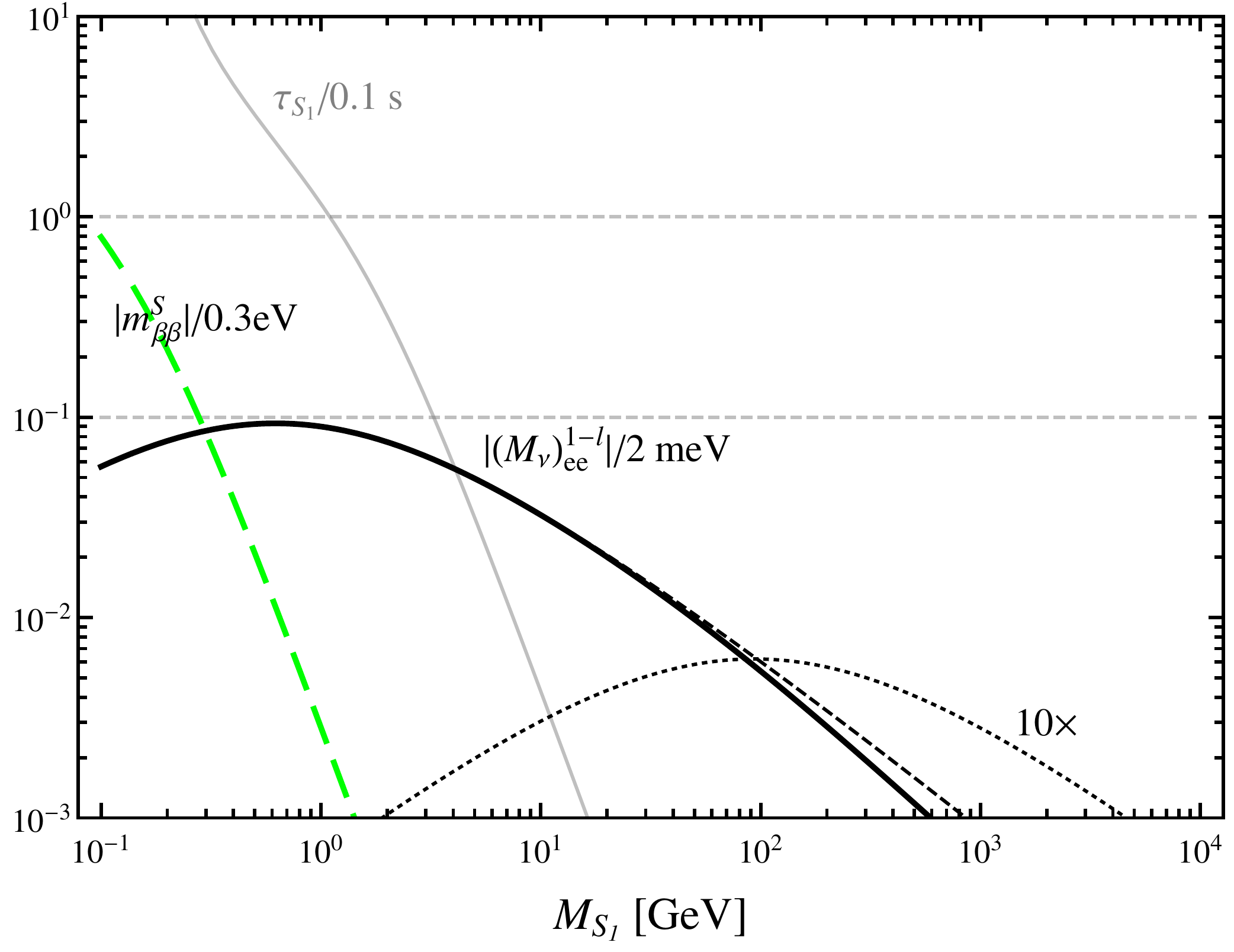}
\caption{\label{fig:loop}
Ratio of one-loop to tree-level contribution (2\,meV) to the $ee$ entry ($\nuless$ 
parameter) of the light neutrino mass matrix 
(solid black) as a function of $M_{S_1}$ ($M_{N_1}$ scales with $M_{S_1}$ 
through $M_{N_1}=R_{N_1}M_{S_1}$).
We also show the contribution coming only from $N_1$ (dashed) and $S_1$ (dotted)
exchange; the latter is multiplied by 10 for visualization.
The contribution for $\nuless$ parameter from $S_1$ exchange (green 
dashed) relative to the limit $0.3$\,eV is shown as well;
we use the expression in \eqref{m.bb.gen}.
The solid gray curve shows the lifetime for $S_1$ relative to 0.1\,s.
The other parameters are fixed as $\md1=7\times 10^{-5}\,\text{GeV}$ and
$R_{N_1}=150$.
}
\end{figure}

As an example, the following values pass all the constraints and are also 
marked in Fig.\,\ref{fig:MDMN} by crosses:
\eqali{
\label{bench}
M_{D_{11}}&=7\times 10^{-5}\,\text{GeV}\,,\quad
M_{N_1}=200\,\text{GeV}\,,\quad
M_{S_1}=1.33\,\unit{GeV}\,,\quad
\bar{\mu}_i\sim 100\,\unit{MeV}\,,
\cr
M_{D_{22}}&=2.1\times 10^{-4}\,\text{GeV}\,,\quad
M_{N_2}=100\,\text{GeV}\,,\quad
M_{S_2}=1\,\unit{GeV}.\quad
}
The intermediate scales $\sqrt{3}u_0f_1'\approx 16\,\unit{GeV}$ and 
$\sqrt{3}u_0f_2'=10\,\unit{GeV}$ can be obtained 
from \eqref{trade:LL}. They set a lower bound for the scales 
$\aver{\eta_i}\sim u_0\gtrsim 10\,\text{GeV}$ and 
$\aver{\varphi_0}\sim\aver{\varphi_1}\gtrsim \mu_i\sim 0.1\,\text{GeV}$ while 
the masses can be chosen $M_\eta\sim u_0\gtrsim M_\varphi \sim 10\,\text{GeV}$.
Using the values in \eqref{bench} as a benchmark, we plot in Fig.\,\ref{fig:loop}
the ratio of the one-loop contribution to the tree-level value of
$|\mbbeta[\nu]|=|(M_\nu^{\rm tree})_{ee}^*|=2\,\text{meV}$ where now we vary $\Ms1$ 
and rescale $M_{N_1}$ simultaneously by fixing $R_{N_1}=150$.
For the benchmark values \eqref{bench}, the one-loop contribution is indeed less 
than 10\% of the tree-level value.
We also show the ratio of the lifetime to the limit of 0.1\,s (solid gray) and 
confirm that $\Ms1$ needs to be larger than around 1\,\text{GeV}.

Finally, we can estimate the amount of cancellation that is built-in in our ESS 
mechanism implementation. Rewriting \eqref{ai:mag:2} in the form of the naive 
seesaw relation,
\eq{
|(M^{\rm tree}_{\nu})_{ee}|=\eps_{ee}\frac{\md1^2}{\Ms1}\,,
}
we extract
\eq{
\eps_{ee}=\frac{\bar{\mu}_i}{M_{N_1}}\approx 10^{-3}\text{ -- }10^{-4}\,,
}
if we use Table \ref{table}.
Analogously, for the $\mu\tau$ entry, we obtain $\eps_{\mu\tau}\approx 
10^{-3}\text{ -- }2\times 10^{-6}$. 
These values are in agreement with the radiative stability conditions discussed in 
Ref.\,\cite{vissani} that estimated a lower bound of $\eps>10^{-6}$ for a GeV-scale 
right-handed neutrino mass.

\section{Other phenomenological constraints and $\Umutau$ breaking}
\label{sec:pheno}

We analyze here other phenomenological constraints coming from the existence of 
GeV-scale heavy neutrino $S_i$ with mixing with the light neutrinos at the order of
\eqali{
\label{mixing:value}
|(\theta V_S)_{e1}|^2 &= 
\left(\frac{\md1}{\sqrt{3}u_0f_1'}\right)^{2}
=\frac{\md1^2}{\Ms1 M_{N_1}}
\cr
&=10^{-12}\times
\left(\frac{\md1 }{10\,\rm keV}\right)^{2}
\left(\frac{100\,\text{GeV}}{M_{N_1}}\right)
\left(\frac{1\,\text{GeV}}{\Ms1}\right)\,.
}
where we have used \eqref{trade:LL} and simplified the notation for $\theta_{\nu 
S}V_S$.
For the values \eqref{bench}, 
\eq{
|(\theta V_S)_{e1}|^2=\frac{\md1^2}{\Ms1 M_{N_1}}
\sim\frac{(7\times 10^{-5}\,\unit{GeV})^2}{200\,\unit{GeV}\times 1.33\,\unit{GeV}}
\sim 2\times 10^{-11}\,.
}
The other mixing angles are either of the same order or vanishing in the 
limit of $\Umutau$ conservation; cf.\,\eqref{nu-S}.
At the same time, the Yukawa couplings to the RHN in our model are even more 
suppressed, 
\eqali{
f_1&\sim \frac{7\times 10^{-5}\,\unit{GeV}}{174\,\unit{GeV}}\sim 4\times 10^{-7}\,,
\cr
f_{2,3}&\sim \frac{2.1\times 10^{-4}\,\unit{GeV}}{174\,\unit{GeV}}\sim 
10^{-6}\,.
}
They are smaller than the electron Yukawa coupling and thus the 
Higgs couplings to the RHN are very much suppressed (their are smaller than 
the mixing $\theta_{\nu S}$).
Hence, the main interactions of the RHN to the SM fields occur through 
active-sterile mixing in \eqref{mixing:value}.

However, it is clear that indirect detection constraints such as lepton 
universality violation or electroweak precision tests are not able to restrict or 
probe such a small mixing angles\,\cite{deppisch,gouvea}.
They are also unobservable through direct detection in 
meson decays \,\cite{deppisch,gouvea,shapo} or in colliders\,\cite{aguila}.
Note that this scenario contrasts with models where Higgsses charged under $\ZZ_3$ 
[or $\Umutau$] may induce large lepton-flavor-violating 
Higgs decays\,\cite{lft:higgs}.

For the same reason, lepton flavor violation (LFV) constraints are very weak in 
our model. The suppression is even larger because LFV processes such as $\mu\to 
e\gamma$ or $\mu\to eee$ are forbidden in the limit of $\Umutau$ conservation. One 
can also see this in 
\eqref{nu-S} as $(\theta V_S)_{ei}(\theta V_S)^*_{\mu i}$ always vanish. 
Being a larger group, $\Umutau$ is more constraining than lepton flavor 
triality\,\cite{lep.triality} and the former only allows $\tau^-\to \mu^+e^-e^-$.
However, when this process is mediated only by heavy neutrinos, it occurs through 
box diagrams that are very much suppressed\,\cite{pilaftsis:lfv}.
These conclusions are not modified when $\Umutau$ breaking effects are considered.
See appendix \ref{ap:dev}.

At last, we can analyze the limits coming from neutrinoless double beta decay, 
which are the strongest involving the mixing with the electron flavor.
Since the active-sterile mixings are all vanishing or of the same order in the 
$\Umutau$ symmetry limit, cf. \eqref{nu-S}, we expect that this process will pose 
the strongest constraint on the mixings.

The half-life of the process is proportional to\,\cite{deppisch,vissani}
\eq{
    \label{half-life}
\frac{1}{T^{0\nu}_{1/2}}\propto 
\left|\frac{\mbbeta[\nu]}{\aver{p^2}}
+ \sum_{i=1}^{n_s}\frac{(\theta V)_{ek}^2M_i}{\aver{p^2}-M^2_i}
\right|^2\,,
}
where $\aver{p^2}\sim -(200\,\text{MeV})^2$ quantifies the effective momentum 
transfer inside the nucleus and $M_i$ represent the masses of the additional heavy 
neutrino states that mix with the three active ones.
The light neutrino contribution depends on
\eq{
\mbbeta[\nu]\equiv \sum_i U_{ei}^2m_i\,,
}
with contributions arising from tree and loop contributions
\eq{
(\mbbeta[\nu])^* =(M_\nu)_{ee}=(M_\nu^{\rm 
tree})_{ee}+(M_\nu^{\onel})_{ee}+\cdots\,.
}
For $\cpmutau$ symmetric theories, it is confined to bands depending on the CP 
parities of the light neutrinos\,\cite{cp.mutau}.
For a review on generic aspects of $\nuless$ see Ref.\,\cite{nuless:rev}.
We are assuming we are confined to the parameter space where the one-loop 
contributions are negligible compared to the tree-level one.

Considering \eqref{half-life}, we can define, in analogy to the light neutrino 
contribution\,\cite{lopez-pavon},
\eq{
\label{m.bb.gen}
\mbbeta[S]\equiv |\aver{p^2}|\sum_{i=1}^{3}\frac{(\theta 
V_S)^2_{ei}M_{S_i}}{M_{S_i}^2+|\aver{p^2}|}\,.
}
where $|\aver{p^2}|\approx (253\,\unit{MeV})^2$
(corresponding to $0.079\times (0.9\,\unit{GeV})^2$ in Ref.\,\cite{petcov.15}) 
and we have already specialized to ${}^{76}\mathrm{Ge}$.
We disregard the subdominant contribution from the heavier states $N_i$.
If the heavy neutrino masses are much larger than the typical momentum transfer in 
the nucleus, $M_i\gg 200\text{ MeV}$, we can approximate
\eq{
\label{nuless}
\mbbeta[S]=|\aver{p^2}| \sum_{i=1}^{3}\frac{(\theta V)^2_{ei}}{\Ms{i}}\,.
}

Taking the GERDA+Helderberg-Moscow limit, 
$T^{0\nu}_{1/2}({}^{76}\mathrm{Ge})\ge 3\times 
10^{25}\rm yr$ at 90\% C.L. \cite{gerda}, it translates into
\eq{
  \label{gerda}
\left|\mbbeta[\nu]+\mbbeta[S]\right|
\lesssim 0.3\,\text{eV}\,.
}
We can see that the contribution from light neutrinos predicted in our model 
\eqref{nuless:0} is at least two orders of magnitude smaller than the limit above.
It remains to be checked if the contribution from $S_i$ exchange can give a larger 
contribution.

In the limit where $\Umutau$ (or LFT) in \eqref{mu-tau} is conserved, only $S_1$ 
couples to the $e$ flavor and thus to $\nuless$; cf. \eqref{theta.ei}.
We can then write
\eq{
\mbbeta[S_1]=-2.67\times 10^{-5}\,\unit{eV}\times
\left(\frac{\md1}{10\,\unit{keV}}\right)^2\frac{10^2}{R_{N_1 } }
\left(\frac{\unit{GeV}}{M_{S_1}}\right)^3\,,
}
where we have assumed that $\Ms1\gg 200\,\unit{MeV}$.
For the values \eqref{bench}, this contribution is negligible.
One could lower the $M_{S_1}$ mass to increase this contribution (including the 
correction in \eqref{m.bb.gen}) but it hits the BBN constraint rather quickly.
Such a feature is illustrated in Fig.\,\ref{fig:loop} where the ratio of 
the $\nuless$ contribution from $S_1$ exchange to the limit of 0.3 eV is shown in 
dashed green. Note that we use the expression \eqref{m.bb.gen} to account for 
$\Ms1<100\,\rm MeV$.
We can see that the $\mbbeta[S]$ is negligible for $\Ms1$ larger than 1.33 GeV.
Even if we allow the lifetime of $S_1$ to be around 1\,s, it will still be
unobservable in future $\nuless$ experiments.
It is possible, however, that $\mbbeta[S]\sim 30\,\rm meV$ for $\Ms1\sim 
300\,\unit{MeV}$ and much larger than the light neutrino contribution.

\section{Conclusions}
\label{sec:concl}

We have presented a new CP symmetry applicable to models with $A_4$ flavor 
symmetry and other groups with the structure $H\rtimes\ZZ_3$ such as $\Delta(27)$. 
To implement this type of CP symmetry, the singlets $\bs{1}'$ that are fermions or 
carry other quantum numbers should appear \textit{in pair} with another $\bs{1}''$ 
with the remaining quantum numbers identical to those of $\bs{1}'$.
This new CP symmetry allows us to avoid the vev alignment problem in close analogy 
to the construction using $\mathtt{L}_\mu-\mathtt{L}_\tau$ and $\cpmutau$ 
symmetries \cite{cp.mutau}.
This feature partly follows because the SM lepton fields are singlets of $A_4$ 
and only feel the $\ZZ_3$ subgroup which is contained in 
$\mathtt{L}_\mu-\mathtt{L}_\tau$.

We have constructed an explicit renormalizable model that leads to a new form for 
the light neutrino mass matrix, cf. \eqref{Mnu:a4cp}.
It retains the successful predictions of $\cpmutau$ -- namely maximal 
$\theta_{23}$, maximal Dirac CP phase and trivial Majorana phases -- but because of 
the $A_4$ structure it also predicts normal hierarchy with the lightest neutrino of 
only few meV; see \eqref{m1:range}.
The CP parities are also restricted to two possibilities which effectively fix 
the effective parameter $\mbbeta$ contributing to neutrinoless double beta 
decay.

The model itself is based on the extended seesaw mechanism which naturally leads to 
relatively light right-handed neutrinos $S_i$ and heavier $N_i$.
After enforcing negligible one-loop contributions to light neutrino masses, ensure 
the ESS approximation and require fast enough decay rate of the BSM states to avoid 
BBN constraints we only find a small allowed region in the parameter space: 
$N_i$ neutrinos lie at the electroweak scale and the lighter $S_i$ lie at the GeV 
scale; see Fig.\,\ref{fig:MDMN}.
To suppress the one-loop contributions, it is required that their Yukawa 
interactions with the SM fields should be smaller than the electron Yukawa 
coupling. Consequently the active-sterile mixing is largely suppressed, rendering 
the right-handed neutrinos practically unobservable in terrestrial experiments.

The flavor structure of the model is largely determined by the approximate 
conservation of the combination $\mathtt{L}_\mu-\mathtt{L}_\tau$ of lepton flavors, 
which suppresses various flavor changing processes such as $\mu\to e\gamma$.
Moreover, only $S_1$ mixes appreciably to 
$\nu_e$ and the mixing of $S_{2,3}$ to the $\mu\tau$ flavors are of the same 
order of magnitude. 

\acknowledgements

The author would like to thank the Maryland Center for Fundamental Physics of 
the University of Maryland at College Park, USA, for the hospitality where this 
work initiated.
The author also thanks Rabindra Mohapatra for discussion on several points.
Partial support by Brazilian Fapesp grant 2013/26371-5 and 2013/22079-8 is 
acknowledged.

\appendix
\section{Block diagonalization of $M_R$}
\label{ap:block}

The ESS mechanism naturally leads to two disparate scales for the right-handed 
neutrinos:  the lighter $M_S$ ($S_{iR}$) and the heavier $M_N$ ($N_{iR}$).
So it is useful to write the complete neutrino mass matrix \eqref{MM} in a basis 
where $M_R$ is block diagonal:
\eq{
\mathbb{M}'=
\mtrx{0 & M_D'^{\tp} \cr M_D' & M_R'}
\approx 
\mtrx{0 & -m_DM_N^{-1}\Lambda^\tp & m_D^\tp\cr
(-m_DM_N^{-1}\Lambda^\tp)^\tp & M_S & 0 \cr
m_D & 0 & M_N'\cr
}\,.
}
The mass matrix $M_S$ is given by \eqref{ess:m}.
The subleading correction to $M_N$ is
\eq{
M_N'=M_N+\ums{2}(\Lambda^\tp\Lambda^*M_N^{*-1}+\tr\!.)\,,
}
where $\tr\!.$ indicates the transpose of the previous matrix.

The block diagonalization is performed by
\eq{
U_R\approx\mtrx{0 & \id\cr \id & 0}
    \mtrx{\id -\theta_R\theta_R^\dag/2 & \theta_R \cr
    -\theta_R^\dag & \id - \theta_R^\dag\theta_R/2}\,,
}
with
\eq{
\label{S-N}
\theta_R^*=\Lambda\, M_N^{-1}\,.
}

Further block diagonalization leads to the results in \eqref{ess:m} and 
\eqref{nu-SN}. 
The complete diagonalization is performed by
\eqali{
\nu_i&\to (\pmns)_{ij}\nu_{jL}+(\theta_{\nu S} V_S)_{ij}S_{jR}^c
+(\theta_{\nu N} V_N)_{ij}N_{jR}^c\,,
\cr
S_{iR}^c&\to (V_S)_{ij}S_{jR}^c
+(-\theta_{\nu S}^\dag\pmns)_{ij}\nu_{iL}+(\theta_RV_N)_{ij}N_{jR}^c\,,
\cr
N_{iR}^c&\to (V_N)_{ij}N_{jR}^c
+(m_D^{\tp -1} M_\nu\pmns)_{ij}\nu_{iL}+(-\theta_R^\dag 
V_S)_{ij}S_{jR}^c\,.
}
The fields on the left-hand side are in the flavor basis and appear in 
\eqref{lag:ESS}; the ones on the right-hand side are the mass eigenfields and 
$\pmns$ is the PMNS matrix in the flavor basis.
We have neglected nonunitary effects and the small mixing angles $\theta_{\nu 
S},\theta_{\nu N},\theta_R$ were already given in Eqs.\,\eqref{nu-SN} and 
\eqref{S-N}.

\section{Comments on the potential}
\label{ap:pot}

Here we justify the approximate conservation of $\Umutau$ that follows from the 
$\ZZ_3$ conserving vevs for $\eta$ in \eqref{111}.

We start by observing that when the potential for $\eta$ is invariant by global 
rephasing, the potential is identical to a potential with three Higgs doublets with 
$A_4$ symmetry and we know that \eqref{111} can be exactly a global 
minimum\,\cite{3hdm}.

The addition of the two independent quartic terms that breaks $U(1)$ but conserves 
$\ZZ_4^D$,
\eq{
I_1=\eta_1^4+\eta_2^4+\eta_3^4
\quad \text{and}\quad
I_2=(\eta_1\eta_2)^2+(\eta_2\eta_3)^2+(\eta_3\eta_1)^2\,,
}
can be chosen to maintain such alignment and also to make $u_0$ real and 
positive.
We stress that these and other quartic terms are not invariant by $\Umutau$ but 
only the $\ZZ_3$ subgroup.
These terms also help to maintain the deviations of $\aver{\eta}$ in the real 
direction since the coefficients are real because of $\cpmutau$.

Now we add the interactions of $\eta$ with $\varphi_0$ and $\varphi_1$.
The relevant terms are
\eq{
\label{eta.phi0}
V\supset \ums{2}M^2_{0}\varphi_0^2+\mu_{0\eta}\varphi_0(\eta^\tp 
\eta+h.c.)\,,
}
and
\eq{
    \label{eta.phi1}
V\supset M_1^2|\varphi_1|^2+
\Big\{\mu_{1\eta}[(\eta\eta)_{1''}+(\eta\eta)_{1'}^*]\,\varphi_1+h.c.\Big\}
\,,
}
where $\mu_{1\eta}$ can be complex and the singlet combination was defined in 
\eqref{singlets}. 
Clearly there is no $U(1)$ rephasing symmetry for $\varphi_1$ and no Goldstone will 
be generated.

The mild hierarchy of ESS scales
\eq{
\text{100\,MeV}\sim \mu_i\ll 
f_i'\sqrt{3}u_0\sim \text{10\,GeV}\,,
}
implies a mild hierarchy between $u_0\sim \aver{\eta_i}$ and 
$\aver{\varphi_0},\aver{\varphi_1}$.
We can choose $f_i'\sim 0.1$ so that $u_0\sim 100\,\text{GeV}\sim M_N$.
For an order one $k_0$, the small $\aver{\varphi_0}\sim 100\,\unit{MeV}$ can be 
generated from \eqref{eta.phi0} by a vev seesaw analogous to type-II 
seesaw\,\cite{type-II}.
In this case $M_{\varphi_0}\sim u_0$ is electroweak scale.
For $\aver{\varphi_1}$ a vev seesaw cannot be implemented because 
$(\eta\eta)_{1''}$  vanishes for the minimum \eqref{111}.
But we can always take $k_1\sim 10^{-2}$, adjust the potential parameters to obtain
$\aver{\varphi_1}\sim 10\,\unit{GeV}$ and make $\mu_{1\eta}$ in \eqref{eta.phi1} 
small enough so that \eqref{111} is only slightly disturbed.
The mass of the lightest physical states of $\varphi_1$ will be around 
$\aver{\varphi_1}$ and heavier than $S_i$.
Note that $k_0\aver{\phi_0}$ and $k_1\aver{\phi_1}$ should be comparable because 
they lead to $\mu_i$.

At last, in principle the new scalars could be produced in Higgs decays through the 
Higgs portal but the current limits on the invisible Higgs decays are still 
weak\,\cite{atlas:inv} and can be avoided by decreasing the portal interactions.

\section{Decay rates for $S_i$}
\label{ap:decay}

In our theory the RHN heavy states $S_i$ are the lightest new states beyond the SM 
which lies at the GeV-scale.
The dominant decay channels involve $Z$ or $W$ exchange through mixing with light 
neutrinos or charged leptons\,\cite{shapo}.
The decays $S_i^c\to S_j^c+\cdots$ are highly suppressed.

To ensure that the production of light nuclear elements in the early Universe 
(Big Bang nucleosinthesis) are not disturbed by the presence of new 
particles, we require that the lifetimes of the new states are shorter than 0.1 
second. In that case these new particles are thermalized much before the BBN era 
and they decay fast enough.
RHNs lighter than around 100 MeV conflict with direct detection constraints and are 
excluded\,\cite{nu:BBN,mnu>mpi}.

Assuming the $\Umutau$ symmetry, the active-sterile mixing \eqref{nu-S} leads to 
the dominant decay channels\,\cite{shapo}
\eqali{
S_1^c&\to \pi^0\nu_e,\,\nu_e\bar{\nu}\nu,\, \pi^+e^-,
\cr
S_{2,3}^c&\to \pi^0\nu_{\mu,\tau},\,\nu_{\mu,\tau}\bar{\nu}\nu,\, \pi^+\mu^-\,.
}
We neglect the decay to other channels.
The decay rates for these processes can be taken from Ref.\,\cite{shapo}:
\eqali{
\Gamma(S_1^c\to \pi^0\nu_e)&=
    |(\theta V_S)_{e1}|^2\frac{G_F^2 f_\pi^2 M^3}{32 \pi }
    \left(1-\frac{m_{\pi^0}^2}{M^2}\right)^2\,,\cr
\Gamma(S_{2}^c\to \pi^0\nu_{\mu+\tau})&=
    \Big(|(\theta V_S)_{\mu 2}|^2+|(\theta V_S)_{\tau 2}|^2\Big)
    \frac{G_F^2 f_\pi^2 M^3}{32 \pi }
    \left(1-\frac{m_{\pi^0}^2}{M^2}\right)^2\,,\cr
\Gamma(S_1^c\to \nu_e\bar{\nu}\nu)&=
    |(\theta V_S)_{e1}|^2\frac{G_F^2 M^5}{192\pi^3}\,,\cr
\Gamma(S_2^c\to \nu_{\mu+\tau}\bar{\nu}\nu)&=
    \Big(|(\theta V_S)_{\mu 2}|^2+|(\theta V_S)_{\tau 2}|^2\Big)
    \frac{G_F^2 M^5}{192\pi^3}\,,\cr
\Gamma(S_1^c\to \pi^+e^-)&=
    |(\theta V_S)_{e1}|^2
    \frac{G_F^2 f_\pi^2|V_{ud}|^2 M^3}{16 \pi }
    \left(
    \left(1-\frac{m_{e}^2}{M^2}\right)^2-
    \frac{m_{\pi^+}^2}{M^2}\left(1+\frac{m_{e}^2}{M^2}\right)\right)
  \cr&\times
    \sqrt{\left(1-\frac{(m_{\pi^+}-m_e)^2}{M^2}\right)
    \left(1-\frac{(m_{\pi^+}+m_e)^2}{M^2}\right) }
    \,,\cr
\Gamma(S_2^c\to \pi^+\mu^-)&=
    |(\theta V_S)_{\mu 2}|^2
    \frac{G_F^2 f_\pi^2|V_{ud}|^2 M^3}{16 \pi }
    \left(
    \left(1-\frac{m_{\mu}^2}{M^2}\right)^2-
    \frac{m_{\pi^+}^2}{M^2}\left(1+\frac{m_{\mu}^2}{M^2}\right)\right)
  \cr&\times
    \sqrt{\left(1-\frac{(m_{\pi^+}-m_\mu)^2}{M^2}\right)
    \left(1-\frac{(m_{\pi^+}+m_\mu)^2}{M^2}\right) }
    \,.
}
In each expression, $M$ refers to the mass of the decaying particle and each 
decay rate contributes twice due to the charge conjugate mode.
Moreover, the expression for $S_3^c$ are identical to the expressions for $S_2^c$ 
and note that we can write
\eqali{
|(\theta V_S)_{e1}|^2&=\frac{\md1 ^2}{M_{S_1}M_{N_1}}\,,\cr
|(\theta V_S)_{\mu 2}|^2+|(\theta V_S)_{\tau 2}|^2
&=\frac{\md2^2}{M_{S_2}M_{N_2}}
\cr
&=2|(\theta V_S)_{\mu 2}|^2=2|(\theta V_S)_{\tau 2}|^2\,.
}
We are also assuming that $\Umutau$ is slightly broken so that $S_{2,3}$ are 
distinct Majorana fermions.
In the exact $\Umutau$ limit, $(S_2^c+iS_3^c+S_2+iS_3)/\sqrt{2}=S_{\mu\bar{\tau}}$ 
forms a Dirac heavy neutrino with $\Umutau$ charge unity while its conjugate 
carries charge $-1$. 
In this case, the decay rates of $S_{\mu\bar{\tau}}$ are the same as $S_2^c$
without the factor two multiplication (the last one would be doubled due to 
diagonal mixing).

\section{Deviations of $\Umutau$}
\label{ap:dev}

In the fermion sector our model is approximately invariant by $\Umutau$, which 
includes $\ZZ_3$ \eqref{triality} of $A_4$.
In the first approximation considered $\Umutau$ is only broken in the neutrino 
sector by small $\mu_i\sim 100\,\text{MeV}$ in \eqref{MM}.
Identical $\Umutau$ charges \eqref{mu-tau} can be assigned to all the lepton fields 
\eqref{mu-tau}
if we change basis to 
\eq{
\label{S'}
S_{iR}^c=(U_\om)_{ij}S_{jR}'^c.
}
We show below the form of the mass matrices in this basis with small
$\Umutau$ breaking.

An additional $\Umutau$ (and also LFT) breaking effect in the neutrino sector is 
induced by deviations in $\aver{\eta}$ from \eqref{111}, which can be parametrized 
as
\eq{
\label{eta:Z3-break}
\aver{\eta}=u_0\Big\{(1,1,1)+\eps_2(1,\om^2,\om)+\eps_3(1,\om,\om^2)\Big\}\,.
}
The deviation is quantified by $|\eps_i|\ll 1$.
$\cpmutau$ is expected to be conserved as there is no CP violating 
interactions for $\eta$. Hence we expect $\eps_3=\eps_2^*$.

The mass matrices \eqref{m.matrix} in the $S_i'$ basis read 
\eqali{
\Lambda'&=\sqrt{3}u_0
\mtrx{1&\eps_3&\eps_2\cr \eps_2&1&\eps_3\cr\eps_3&\eps_2&1}
\diag(f_i')\,,
\cr
\mu'&=
\ums{3}\mtrx{
\mu_1+\mu_2+\mu_3 &\mu_1+\om\mu_2+\om^2\mu_3&\mu_1+\om^2\mu_2+\om\mu_3\cr
\star &\mu_1+\om^2\mu_2+\om\mu_3&\mu_1+\mu_2+\mu_3\cr
\star &\star&\mu_1+\om\mu_2+\om^2\mu_3\cr
}
}
where the $\Umutau$ breaking parametrization \eqref{eta:Z3-break} for 
$\aver{\eta}$ was used.
The explicit change of basis is induced by
\eq{
\Lambda'=U_\om \Lambda,\quad
\mu'=U_\om\mu U_\om\,.
}
Conservation of $\cpmutau$ implies $\eps_3=\eps_2^*$ and real $\mu_i$.
In the mass matrices it implies the usual $\cpmutau$ invariance:
\eq{
X^\tp\Lambda'X=\Lambda'^*\,,\quad
X^\tp\mu'X=\mu'^*\,.
}

In the same basis, the $S_i$ neutrino mass matrix \eqref{MS:1} becomes 
\eq{
\label{MS'}
-M_{S}'^{(0)}=
\mtrx{\Msz[1] &&\cr
&&\Msz[2]\cr
&\Msz[2]&\cr}\,,
}
where $\Msz[i]$ were given in \eqref{masses:S} and we have added the superscript 
$^{(0)}$ to indicate the $\Umutau$ limit explicitly. 
A generic deviation respecting $\cpmutau$ arising from $\aver{\eta}$ can be 
parametrized by
\eq{
    \label{delta:MS}
-\delta M_{S}'=\Msz[1]
\mtrx{0&z_{12}&z_{13}\cr
\star &z_{22}& 0\cr
\star & \star&z_{33} }\,,
}
where $z_{13}=z_{12}^*$, $z_{33}=z_{22}^*$.

The combination $M_S'=M_S'^{(0)}+\delta M_{S}'$ is now diagonalized by 
\eq{
V_S'= iU_{23}\cO_\eps\,,
}
where $U_{23}$ denotes the maximal mixing matrix in \eqref{U23}.
One can check that $\cO_\eps$ is a real orthogonal matrix given by
\eq{
    \label{Oe}
\cO_\eps\approx
\mtrx{1&-d_1'&-d_2'\cr d_1&c_\theta&-s_\theta\cr d_2&s_\theta&c_\theta}\,.
}
The small parameters $d_i$ are combinations of the small quantities in 
\eqref{delta:MS} and are defined by
\eq{
-U_{23}^\tp \delta M_S'U_{23}=
\Msz[1]\mtrx{0&d_1&d_2\cr d_1&c_1&c_2\cr d_2&c_2&-c_1}\,;
}
all $d_i,c_i$ are real.
The primed $d_i'$ are rotated as 
\eq{
\mtrx{d_1'\cr d_2'}
=
\mtrx{
c_\theta& s_\theta\cr -s_\theta&c_\theta}
\mtrx{d_1\cr d_2}\,,
}
with angle $\tan 2\theta=c_2/c_1$.
One can note that the angle $\theta$ depends only on the deviation parameters $c_i$ 
and does not need to be small due to the degeneracy $\Msz[2]=\Msz[3]$.
The formula \eqref{Oe} is valid as long as $d_i,c_i\ll 1$ and covers the case where 
$\Msz[1]\gg \Msz[2]\sim \Msz[1]d_i\sim \Msz[1]c_i$ so that the mass splitting for 
$S_{2,3}$ can be substantial:
\eqali{
M_{S_2}&=\Msz[2]+\Msz[1]\sqrt{c_1^2+c_2^2}\,,\cr
M_{S_3}&=\Msz[2]-\Msz[1]\sqrt{c_1^2+c_2^2}\,.
}
We are adopting $M_{S_3}<M_{S_2}$.

Putting all together we find the deviation from \eqref{nu-S}:
\eqali{
    \label{nu-S:dev}
(\theta V_S)&\approx |(\theta V_S^{(0)})_{e1}|
\diag(1,k,k)
\mtrx{1&-\eps_3^*&-\eps_2^*\cr -\eps_2^*&1&-\eps_3^*\cr-\eps_3^*&-\eps_2^*&1}
\times iU_{23} \cO_\eps
\,,\cr
&=
(\theta V_S^{(0)})_{e1} \diag(1,k,k)\times
\left(
\begin{array}{ccc}
 1 & \eps_{12} & \eps_{13} \\
 \eps_{21} & x_{22} & x_{23} \\
 \eps_{31} & x_{32} & x_{33} \\
\end{array}
\right)
\,,
}
where $\eps_{ij}$ are small parameters that depend on the small parameters 
$\eps_{2,3}$ while $x_{ij}$ are order one, approximately unitary, 
quantities. The deviation from maximal (23) mixing in \eqref{nu-S} can be large due 
to $S_{2,3}$ mass degeneracy in the $\Umutau$ limit.
Again the superscript $(0)$ denotes the $\Umutau$ limit.
Note that $(\theta V_S)$ has the structure
\eq{
\mtrx{u_1&u_2&u_3\cr w_1&w_2&w_3\cr w_1^*&w_2^*&w_3^*}\,,
}
characteristic of $\cpmutau$ invariance\,\cite{cp.mutau,real.sym}

Considering the deviation \eqref{nu-S:dev} in $\nu-S$ mixing, we can include the 
effects of $M_{S_{2,3}}$ exchange in $\nuless$ as
\eq{
\mbbeta[S]=\mbbeta[S_1]\left\{
   1+\eps_{12}^2\frac{\Msz[1]}{M_{S_2}}+\eps_{13}^2\frac{\Msz[1]}{M_{S_3}}
\right\}\,,
}
where $M_{S_{2,3}}$ are now nondegenerate and include the $\Umutau$ breaking 
effects.
It is clear that the contribution of $S_2$ ($S_3$) exchange can be comparable to 
$S_1$ exchange only if 
\eq{
M_{S_1}/M_{S_{2,3}}\sim O(1/\eps^2)\,.
}
This cannot happen in our theory.

We can also confirm that $\Umutau$ breaking is not enough to induce observable 
lepton flavor violating processes such as $\mu\to e\gamma$.
The vanishing rate is now proportional to the $\Umutau$ breaking effects.
Considering only $S_i$ in the loop, the branching ratio yields\,\cite{ibarra.petcov}
\eq{
B(\mu\to e\gamma)\sim 2\times 10^{-30}\times 
\left|\frac{\eps_{21}}{0.1}\right|^2
\times\left|\frac{(\theta V)_{eS_1}}{10^{-6}}\right|^4
\tilde{G}_1^2
\,,
}
where $\tilde{G}_i=G(M_{S_i}^2/M_W^2)-G(0)$ and $G(x)$ is defined in 
Ref.\,\cite{ibarra.petcov}.
For example, $G(1^2/80^2)-G(0)\approx -10^{-4}$.
Therefore, the predicted rate is much below the current MEG limit $B(\mu\to 
e\gamma)<2.4\times 10^{-12}$\,\cite{MEG} and there is no constraint even if 
$(\theta V)_{eS_1}$ is as large as 1\%.
One can also check that $S_{2,3}$ contributions lead to similar results.
Future $\mu\to e$ conversion experiments in nuclei\,\cite{alonso} can improve 
the limit by few orders of magnitude but our model predictions are still suppressed.
Hence, LFV processes constraints are much weaker than $\nuless$ in our 
model.


\end{document}